\documentclass[prx, twocolumn, groupedaddress]{revtex4-1}
\usepackage{graphicx,color}
\usepackage{amssymb}   % for math
\usepackage{amsmath}
\usepackage{amsfonts}
\usepackage{mathdots}
\usepackage{hyperref}
\definecolor{myblue}{RGB}{46, 48,146}
\hypersetup{colorlinks=true,linkcolor=myblue,citecolor=myblue,urlcolor=myblue, linktocpage}
\usepackage{braket}
\usepackage{bm}
\usepackage{natbib}
\usepackage{gensymb}
\usepackage{threeparttable}
\usepackage{multirow}
\usepackage{longtable}
\usepackage{booktabs}
\usepackage{gensymb}

\begin{document}
	\title{Higher-Order Nonlinear Anomalous Hall Effects Induced by Berry Curvature Multipoles}	
	\author{Cheng-Ping Zhang$^1$}
	\thanks{These authors contributed equally to this work}
	\author{Xue-Jian Gao$^1$}
	\thanks{These authors contributed equally to this work}
	\author{Ying-Ming Xie$^1$}	
	\author{Hoi Chun Po$^2$}
	\author{K. T. Law$^1$} \thanks{phlaw@ust.hk}
	\affiliation{$^1$Department of Physics, Hong Kong University of Science and Technology, Clear Water Bay, Hong Kong, China} 
	\affiliation{$^2$Department of Physics, Massachusetts Institute of Technology, Cambridge, Massachusetts, USA} 
	
	\begin{abstract}
		In recent years, it has been shown that Berry curvature monopoles and dipoles play essential roles in the anomalous Hall effect and the nonlinear Hall effect respectively. In this work, we demonstrate that Berry curvature multipoles (the higher moments of Berry curvatures at the Fermi energy) can induce higher-order nonlinear anomalous Hall (NLAH) effect. Specifically, an AC Hall voltage perpendicular to the current direction emerges, where the frequency is an integer multiple of the frequency of the applied current. Importantly, by analyzing the symmetry properties of all the 3D and 2D magnetic point groups, we note that the quadrupole, hexapole and even higher Berry curvature moments can cause the leading-order frequency multiplication in certain materials. To provide concrete examples, we point out that the third-order NLAH voltage can be the leading-order Hall response in certain antiferromagnets due to Berry curvature quadrupoles, and the fourth-order NLAH voltage can be the leading response in the surface states of topological insulators induced by Berry curvature hexapoles. Our results are established by symmetry analysis, effective Hamiltonian and first-principles calculations. Other materials which support the higher-order NLAH effect are further proposed, including 2D antiferromagnets and ferromagnets, Weyl semimetals and twisted bilayer graphene near the quantum anomalous Hall phase.
	\end{abstract}
	
	\maketitle
	
	\emph{Introduction.}---The Hall effect is a fascinating phenomenon that a Hall voltage perpendicular to the applied current direction can be generated under an external magnetic field. In ferromagnets, a Hall voltage can be created in the absence of an external magnetic field, which is known as the anomalous Hall effect. It has an intrinsic contribution from the Berry curvature monopole which is the integral of Berry curvature over occupied states \cite{Nagaosa2010anomalous, Xiao2010berry}. Recently, anomalous Hall effects are also discovered in antiferromagnets \cite{Chen2014anomalous, Kubler2014non-collinear, Surgers2014large, Nakatsuji2015large, Nayak2016large, Suzuki2016large}.
	
	Surprisingly, it was pointed out recently that a Hall voltage can be induced even in  time-reversal invariant systems \cite{Sodemann2015quantum}, in which the generated Hall voltage doubles the frequency of the applied AC electric current. This so-called nonlinear Hall effect is induced by the Berry curvature dipole, which is the first moment of the Berry curvature over occupied states. The nonlinear Hall effect has been observed experimentally in bilayer and multilayer WTe$_2$ \cite{Ma2019observation, Kang2019nonlinear} and more recently in twisted WSe$_2$ \cite{Huang2020giant, Hu2020nonlinear}. However, in principle, higher Berry curvature moments can be non-vanishing and their physical consequences are not known.

	In this work, we provide a general theory for higher-order nonlinear anomalous Hall (NLAH) effects which can be induced by the Berry curvature multipoles such as quadrupole \cite{Parker2019diagrammatic}, hexapole and higher-order multipoles. Specifically, an AC Hall voltage with frequency which is an integer multiple of the frequency of the applied AC current can be generated by Berry curvature multipoles. The higher-order effects are generally expected to be small compared to lower-order effects. However, we point out that Berry curvature quadrupole, hexapole and even higher-order multipoles can cause the leading-order effects when lower-order Berry curvature moments are forced to vanish by crystal symmetry. Magnetic point groups (MPGs) which allow higher-order Berry curvature moments to be the leading-order moments are listed in Table \ref{table:01} and Table \ref{table:02} respectively for three-dimensional (3D) and two-dimensional (2D) materials. To give concrete examples, we point out that in antiferromagnets such as monolayer SrMnBi$_2$, the NLAH effect induced by Berry curvature quadrupole is the leading-order Hall response, as both the anomalous Hall and the nonlinear Hall effect are prohibited by symmetry. Furthermore, with a current easily accessible in experiments, the third-order NLAH voltage can be of the order $\sim 10 \; \mathrm{\mu V}$, which is comparable with the nonlinear Hall voltage in WTe$_2$ \cite{Ma2019observation, Kang2019nonlinear}. We further point out that the surface states of topological insulators with $C_{3v}$ symmetry support fourth-order NLAH effect due to the Berry curvature hexapole which is the lowest non-vanishing moment. 
	
	The rest of the paper is organized as follows. We first use the Boltzmann equation approach to establish the relationship between the AC conductivity and the Berry curvature multipoles. Second, the symmetry properties of the Berry curvature multipoles in all MPGs are analyzed and summarized in Table \ref{table:01} and Table \ref{table:02}. Third, to be specific, we show explicitly how the third-order NLAH effect induced by Berry curvature quadrupole becomes the leading-order Hall response in $4'm'm$ (in Hermann-Mauguin notation~\cite{Notation}) MPG, which is the symmetry for monolayer antiferromagnet SrMnBi$_2$. We further show that the fourth-order NLAH effect induced by Berry curvature hexapole is the lowest order response in $C_{3v}$ point group, and the theory applies to the surface states of topological insulators. The third-order NLAH effect may also be observed in other candidate materials, including 2D antiferromagnets and ferromagnets, Weyl semimetals and twisted bilayer graphene near the quantum anomalous Hall phase.
	%	Since the third-order NLAH effect is a direct evidence of time-reversal symmetry breaking, we also suggest that it can be a distinctive signature to detect 2D antiferromagnetism, which is difficult to detect with other methods \cite{Gibertini2019magnetic}, as the anomalous Hall response and magneto-optical Kerr effect which are commonly used to detect ferromagnetism usually vanish in antiferromagnets.
	
	\emph{Nonlinear conductivity and Berry curvature multipoles}--- In this section, we establish the connection between the nonlinear conductivity and Berry curvature multipoles using the Boltzmann equation approach \cite{Sodemann2015quantum, Parker2019diagrammatic}. We focus only on the intraband contribution, which is valid when the frequency is much lower than the band gaps between adjacent bands. Recall the semiclassical equations of electron motion:
	\begin{eqnarray}
		\frac{d}{dt}\bm{r} &=& \frac{1}{\hbar} \nabla_{\bm{k}}\varepsilon_{\bm{k}} + \frac{e}{\hbar} \bm{E} \times \bm{\Omega},\\
		\frac{d}{dt}\bm{k} &=& -\frac{e\bm{E}}{\hbar},
	\end{eqnarray}
	where $\bm{E}=\bm{E}(t)$ is the time-dependent applied electric field and $\bm{\Omega}$ is the Berry curvature.
	
	The electric current is given by the integral of physical velocity:
	\begin{equation}
		\bm{j}(t)  = -e \int_{\bm{k}} f(\bm{k},t) \frac{d\bm{r}}{dt},	
	\end{equation}
	where $\int_{\bm{k}} = \int d^d k/(2\pi)^d$, and $d$ is the dimensionality. The time evolution of distribution function $f(\bm{k},t)$ is given by the Boltzmann equation:
	\begin{equation}
		\frac{d\bm{k}}{dt} \cdot \nabla_{\bm{k}} f(\bm{k},t) + \partial_{t} f(\bm{k},t) = \frac{f_{0}-f(\bm{k},t)}{\tau},
	\end{equation}
	where $f_0$ is the equilibrium Fermi-Dirac distribution function and $\tau$ represents the relaxation time.
	
	With a harmonic electric field $\bm{E}(t)=\mathrm{Re}\{E_{\alpha}e^{i\omega t}\bm{\hat{e}}_{\alpha}\}$ (Greek letters $\alpha, \beta, \gamma = x, y, z$ represent the spatial indices), the current responses can be obtained order by order (see Appendix~\ref{AppendixB} for details).
	
	The first-order response is at the same frequency as the driving force: $j^{(1)}_{\mu}(t)=\mathrm{Re}\{\sigma^{(1)}_{\mu\alpha}(\omega)E_{\alpha}e^{i\omega t}\}$, with
	\begin{equation}
		\sigma^{(1)}_{\mu\alpha}(\omega)=\frac{e^{2}}{\hbar}\int_{\bm{k}}f_{0}(\frac{\partial_{\mu}\partial_{\alpha}\varepsilon_{\bm{k}}}{\hbar\widetilde{\omega}}-\epsilon_{\mu\alpha\beta}\Omega_{\beta}),
	\end{equation}
	where $\partial_{\alpha} = \partial/\partial_{k_{\alpha}}$ and $\epsilon_{\mu\alpha\beta}$ is the Levi-Civita tensor. $\widetilde{n\omega}$ represents $i n\omega+\gamma$ \cite{OmegaTilde} and $\gamma=1/\tau$. The first term is the usual Drude conductivity, which is symmetric with respect to the two indices: $\sigma^{(1), D}_{\mu\alpha}(\omega) = \sigma^{(1), D}_{\alpha\mu}(\omega)$. The second term is the intrinsic contribution to the anomalous Hall conductivity from the integral of Berry curvature $\int_{\bm{k}}f_{0}\Omega_{\beta}$, which can be viewed as Berry curvature monopole. The anomalous Hall conductivity is defined as the anti-symmetric part of the conductivity tensor: $\sigma^{(1), H}_{\mu\alpha}(\omega) = - \sigma^{(1), H}_{\alpha\mu}(\omega)$, which vanishes when time-reversal symmetry is present, as required by Onsager reciprocal relation \cite{Nagaosa2010anomalous}. As an analogy, we define the NLAH conductivity as the anti-symmetric part of the nonlinear conductivity tensor following Ref. \cite{Nandy2019symmetry}, in order to distinguish it from the Drude-like contributions.
	
	The second-order response consists of a rectified current and a second harmonic generation: $j^{(2)}_{\mu}(t)=\mathrm{Re}\{\sigma^{(2)}_{\mu\alpha\beta}(0)E_{\alpha}E_{\beta}^{*}+\sigma^{(2)}_{\mu\alpha\beta}(2\omega)E_{\alpha}E_{\beta}e^{2i\omega t}\}$, with
	\begin{eqnarray}\nonumber
		\sigma^{(2)}_{\mu\alpha\beta}(0)&=&-\frac{e^{3}}{2\hbar^{3}}\int_{\bm{k}}f_{0}\frac{\partial_{\mu}\partial_{\alpha}\partial_{\beta}\varepsilon_{\bm{k}}}{\gamma\widetilde{\omega}}\\
		&&+\frac{e^{3}}{2\hbar^{2}}\frac{\epsilon_{\mu\alpha\gamma}}{\widetilde{\omega}}D_{\beta\gamma},\\\nonumber
		\sigma^{(2)}_{\mu\alpha\beta}(2\omega)&=&-\frac{e^{3}}{2\hbar^{3}}\int_{\bm{k}}f_{0}\frac{\partial_{\mu}\partial_{\alpha}\partial_{\beta}\varepsilon_{\bm{k}}}{\widetilde{\omega}(\widetilde{2\omega})}\\
		&&+\frac{e^{3}}{2\hbar^{2}}\frac{\epsilon_{\mu\alpha\gamma}}{\widetilde{\omega}}D_{\beta\gamma}.
	\end{eqnarray}
	Each conductivity tensor contains two terms. The first term is the Drude-like contribution and the second term is the nonlinear Hall conductivity induced by Berry curvature dipole $D_{\alpha\beta}=\int_{\bm{k}}f_{0}\partial_{\alpha}\Omega_{\beta}$. The second term is the origin of the nonlinear Hall effect first pointed out by Sodemann and Fu \cite{Sodemann2015quantum} which has attracted many theoretical and experimental studies in recent years \cite{Ma2019observation, Kang2019nonlinear, you2018berry, zhang2018electrically, zhang2018berry, facio2018strongly, du2018band, battilomo2019berry, zhou2020highly, Hu2020nonlinear, Huang2020giant}.
	
	\begin{table*}
		\begin{threeparttable}
			\caption{Leading-order intrinsic anomalous Hall responses in all the 122 3D magnetic point groups}
			\begin{ruledtabular}\label{table:01}
				\begin{tabular}{c c}
					leading-order responses & magnetic point groups \tabularnewline [0.5ex]
					\hline 
					\multirow{2}{*}{No anomalous} & $\bar{1}1'$, $\bar{1}'$, $2/m{1}'$, $2'/m$, $2/m'$, $mmm1'$, $m'mm$, $m'm'm'$, $4/m1'$
					$4/m'$, $4'/m'$, $4/mmm1'$, \tabularnewline
					\multirow{2}{*}{Hall response}& $4/m'mm$, $4'/m'm'm$, $4/m'm'm'$, $\bar{3}1'$, $\bar{3}'$, $\bar{3}m1'$, $\bar{3}'m$, $\bar{3}'m'$, $6/m1'$, $6'/m$, $6/m'$, \tabularnewline
					& $6/mmm1'$, $6/m'mm$, $6'/mmm'$, $6/m'm'm'$, $m\bar{3}1'$, $m'\bar{3}'$, $m\bar{3}m1'$, $m'\bar{3}'m$, $m'\bar{3}'m'$ \tabularnewline\tabularnewline
					\multirow{3}{*}{1st-order} & $1$, $\bar{1}$, $2$, $2'$, $m$, $m'$, $2/m$, $2'/m'$, $2'2'2$, $m'm2'$, $m'm'2$, $m'm'm$, \tabularnewline
					& $4$, $\bar{4}$, $4/m$, $42'2'$, $4m'm'$, $\bar{4}2'm'$, $4/mm'm'$, $3$, $\bar{3}$, $32'$, $3m'$, \tabularnewline
					& $\bar{3}m'$, $6$, $\bar{6}$, $6/m$, $62'2'$, $6m'm'$, $\bar{6}m'2'$, $6/mm'm'$ \tabularnewline \tabularnewline
					\multirow{3}{*}{2nd-order} & $11'$, $21'$, $m1'$, $222$, $2221'$, $mm2$, $mm21'$, $41'$, $4'$, $\bar{4}1'$, $\bar{4}'$, $422$, $4221'$, $4'22'$, \tabularnewline
					& $4mm$, $4mm1'$, $4'm'm$, $\bar{4}2m$, $\bar{4}2m1'$, $\bar{4}2'm$, $\bar{4}2m'$, $31'$, $32$, $321'$, $3m$, $3m1'$, \tabularnewline
					& $61'$, $6'$, $622$, $6221'$, $6'22'$, $6mm$, $6mm1'$, $6'mm'$, $23$, $231'$, $432$, $4321'$, $4'32'$ \tabularnewline \tabularnewline
					\multirow{2}{*}{3rd-order} & $mmm$, $4'/m$, $4/mmm$, $4'/mmm'$, $\bar{3}m$, $\bar{6}'$, $6'/m'$, $\bar{6}m2$, \tabularnewline 
					& $\bar{6}’m’2$, $\bar{6}’m2’$, $6/mmm$, $6’/m’mm’$, $m\bar{3}$, $\bar{4}’3m’$, $m\bar{3}m’$ \tabularnewline \tabularnewline
					4th-order & $\bar{6}1'$, $\bar{6}m21'$, $\bar{4}3m$, $\bar{4}3m1'$ \tabularnewline \tabularnewline
					5th-order & $m\bar{3}m$ \tabularnewline
				\end{tabular}
			\end{ruledtabular}
		\end{threeparttable}
	\end{table*}
	
	In this work, we focus on the higher-order responses which importantly can be the leading-order responses under certain MPG symmetries as detailed in the following sections. The third-order response is composed of currents at both the same and triple the fundamental frequency: $j^{(3)}_{\mu}(t)=\mathrm{Re}\{\sigma^{(3)}_{\mu\alpha\beta\gamma}(\omega)E_{\alpha}E_{\beta}E_{\gamma}^{*}e^{i\omega t}+\sigma^{(3)}_{\mu\alpha\beta\gamma}(3\omega)E_{\alpha}E_{\beta}E_{\gamma}e^{3i\omega t}\}$, with
	\begin{eqnarray}\nonumber
		\sigma^{(3)}_{\mu\alpha\beta\gamma}(\omega)&=&\frac{3e^{4}}{4\hbar^{4}}\int_{\bm{k}}f_{0}\frac{\partial_{\mu}\partial_{\alpha}\partial_{\beta}\partial_{\gamma}\varepsilon_{\bm{k}}}{\widetilde{\omega}(\widetilde{-\omega})(\widetilde{2\omega})}\\
		&&-\frac{e^{4}}{4\hbar^{3}}[\frac{2\epsilon_{\mu\alpha\delta}}{\widetilde{\omega}(\widetilde{-\omega})}Q_{\beta\gamma\delta}+\frac{\epsilon_{\mu\gamma\delta}}{\widetilde{\omega}(\widetilde{2\omega})}Q_{\alpha\beta\delta}],\\\nonumber
		\sigma^{(3)}_{\mu\alpha\beta\gamma}(3\omega)&=&\frac{e^{4}}{4\hbar^{4}}\int_{\bm{k}}f_{0}\frac{\partial_{\mu}\partial_{\alpha}\partial_{\beta}\partial_{\gamma}\varepsilon_{\bm{k}}}{\widetilde{\omega}(\widetilde{2\omega})(\widetilde{3\omega})}\\\label{eq:3rd-harmonic}
		&&- \frac{e^{4}}{4\hbar^{3}}\frac{\epsilon_{\mu\alpha\delta}}{\widetilde{\omega}(\widetilde{2\omega})}Q_{\beta\gamma\delta}.
	\end{eqnarray}
	The first term is the Drude-like contribution and the second term is the NLAH conductivity induced by Berry curvature quadrupole \cite{Parker2019diagrammatic}, which is defined as
	\begin{equation}\label{eq:BCQuadrupole}
		Q_{\alpha\beta\gamma} = \int_{\bm{k}}f_{0}\partial_{\alpha}\partial_{\beta}\Omega_{\gamma}.
	\end{equation}
	It can be generalized to multi-band cases by summing up the contributions from all bands.
	
	The quadrupole can also be rewritten as
	\begin{equation}\label{eq:BCQuadrupole2}
		Q_{\alpha\beta\gamma}=-\int_{\bm{k}}(\partial_{\alpha}\varepsilon_{\bm{k}})(\partial_{\beta}\Omega_{\gamma})f'_{0}(\varepsilon_{\bm{k}}-\mu),
	\end{equation}
	which indicates that the NLAH effect induced by Berry curvature quadrupole is a Fermi liquid property. Similarly, the Berry curvature hexapole is defined as
	\begin{equation}\label{eq:BCHexapole}
		H_{\alpha\beta\gamma\delta} = \int_{\bm{k}}f_{0}\partial_{\alpha}\partial_{\beta}\partial_{\gamma}\Omega_{\delta},
	\end{equation}
	and the higher-order moments can be defined in a similar manner. The higher-order nonlinear conductivity and their relations to higher-order Berry curvature moments can be found in Appendix~\ref{AppendixB}.
	
	\emph{Symmetry analysis of Berry curvature multipoles.}---As shown in the last section, Berry curvatures contribute to the higher-order conductivity in general \cite{Parker2019diagrammatic}. In this section, we analyze the symmetry properties of Berry curvature multipoles. Taking the Berry curvature quadrupole as an example, we point out that time-reversal symmetry forces the Berry curvature quadrupoles to be zero. However, for materials belonging to 66 (out of the 122) MPGs which break time-reversal symmetry, the Berry curvature quadrupole can be finite. Moreover, in 15 MPGs as listed in Table \ref{table:01}, the quadrupole is the lowest order non-vanishing Berry curvature moment. %Even with finite Berry curvature quadrupoles, the higher-order responses are more difficult to detect. In the following, we show that for materials belonging to the 15 MPG symmetries listed in Table \ref{table:01}, the first and the second-order effects are forced zero, and the third-order effect induced by Berry curvature quadrupoles becomes the leading-order effect. These materials provide platforms for the study of Berry curvature multipole effects. %For completeness, we also point out that the fourth-order and the fifth-order response due to Berry curvature octopole and 16-pole can be the leading-order response in materials with certain MPGs as listed in Table \ref{table:01}.
	
	To have finite Berry curvature quadrupole, we note that under time-reversal symmetry $\mathcal{T}$: $\partial_{\alpha} \rightarrow -\partial_{\alpha}$ and $\Omega_{\gamma} \rightarrow -\Omega_{\gamma}$ and therefore,  according to Eq.~\ref{eq:BCQuadrupole}, the Berry curvature quadrupole vanishes.  As a result, only materials which break time-reversal symmetry can have transport responses induced by Berry curvature quadrupoles. Under general spatial symmetries, since the Berry curvature is a pseudovector, the Berry curvature quadrupole transforms as a rank-3 pseudotensor. Therefore a symmetry operation $\Lambda$ imposes constraint on the form of the quadrupole:
	\begin{equation}\label{eq:symmetry}
		Q_{\alpha\beta\gamma} =\pm\mathrm{det}(\Lambda) \Lambda_{\alpha\alpha'}\Lambda_{\beta\beta'}\Lambda_{\gamma\gamma'}Q_{\alpha'\beta'\gamma'},
	\end{equation}
	where +(-) is taken for unitary(antiunitary) operations. Furthermore, the quadrupole is symmetric with respect to the first two indices: $Q_{\alpha\beta\gamma}=Q_{\beta\alpha\gamma}$ as indicated by Eq. \ref{eq:BCQuadrupole}, because the order of derivatives are interchangeable.
	
	From above, we note that the Berry curvature quadrupole transforms exactly the same as piezomagnetic tensor $\chi_{\alpha\beta\gamma}$, which generates a magnetization $M_{\gamma} = \chi_{\alpha\beta\gamma} \mathcal{E}_{\alpha\beta}$ when a strain $\mathcal{E}_{\alpha\beta}$ is applied. Because the magnetization is a pseudovector and strain is a symmetric tensor, the piezomagnetic tensor is also a rank-3 pseudotensor with the first two indices to be symmetric $\chi_{\alpha\beta\gamma}=\chi_{\beta\alpha\gamma}$. Out of the 122 MPGs, 66 of them are piezomagnetic \cite{Newnham2005properties} and therefore support nonzero Berry curvature quadrupoles, whose explicit forms are listed in Table~\ref{table:S01} of Appendix~\ref{AppendixC}.
	
	Among the 66 MPGs with Berry curvature quadrupoles, 31 of them have finite Berry curvature monopoles and 20 of them have Berry curvature dipoles as the lowest order non-vanishing moment. Importantly, as listed in Table \ref{table:01}, there are 15 MPGs in which the Berry curvature quadrupole is the leading non-vanishing moment. %since $\partial_{\alpha}\partial_{\alpha}$ is invariant under all symmetry operations, the trace of quadrupole $Q_{\alpha\alpha\gamma}$ has the same symmetry property as the Berry curvature monopole which is the integral of Berry curvature $\int[d\bm{k}]f_{0}\Omega_{\gamma}$, thus is finite for the 31 MPGs which are compatible with linear anomalous Hall effect as listed in Table \ref{table:01}. %Similarly, all the MPGs compatible with Berry curvature dipole allows non-vanishing Berry curvature octopole.
	
	The same analysis can be applied to the higher-order Berry curvature moments. In general for all the 122 3D MPGs, all the odd-order responses require time-reversal symmetry breaking while all the even-order effects require inversion symmetry breaking. The $n$-th order NLAH effect is contributed by the ($n-1$)-th moment of Berry curvature, which transforms as a rank-$n$ pseudotensor. The details of the transformation properties of the Berry curvature multipoles can be found in the Appendix~\ref{AppendixC}, and the leading-order moments of the Berry curvature (therefore the leading-order intrinsic anomalous Hall responses) are obtained accordingly \cite{Gallego2019automatic}, as listed in Table \ref{table:01}. There are 32 MPGs (out of 122) which respect the combination of inversion and time-reversal symmetries $\mathcal{I}\mathcal{T}$, forcing the Berry curvature to vanish in the entire Brillouin zone. Among the remaining 90 MPGs, 31 of them exhibit anomalous Hall effect, which can have nonzero spontaneous magnetization \cite{Newnham2005properties}.	The 2nd-order NLAH effect is the leading-order Hall response in 39 MPGs and the 3rd-order NLAH effect is the leading response in 15 MPGs. There are 4 MPGs in which the hexapole is the leading-order moment, while in $m\bar{3}m$ MPG the octopole is the leading-order moment.
%	2nd-order NLAH effect is the leading-order Hall response in 39 MPGs and 3rd-order NLAH effect is the leading response in 15 MPGs. There are 4 MPGs in which the 4th-order NLAH effect is the leading-order response, while in $m\bar{3}m$ MPG the 5th-order NLAH effect is the leading response.
	
	Similarly, we also study the leading-order responses for the 31 MPGs in 2D space. The ($n+1$)-th order NLAH effect is contributed by the $n$-th moment of Berry curvature, which in 2D space has $n+1$ independent components: $\int_{\bm{k}}f_{0}(\partial_{x})^{l}(\partial_{y})^{n-l}\Omega$, with $l=0,1,\dots,n$. By linear combination, they can be rearranged as  $\int_{\bm{k}}f_{0}\partial_{+}^{n-l}\partial_{-}^{l}\Omega$ with $\partial_{\pm}=\partial_{x}\pm i\partial_{y}$, which form the eigenvectors of the angular momentum operator, with quantum numbers $\pm n$, $\pm (n-2)$, $\cdots$. Apart from the zero angular momentum components, such as the monopole and the trace of quadrupole, all the other components are forced to vanish under a $p$-fold rotational symmetry with $p > n$. Therefore, if an additional time-reversal or mirror symmetry is present, which forces the zero angular momentum components to vanish, then the leading-order Berry curvature moment under a $p$-fold rotation has the order $n = p$, as shown in Table \ref{table:02}. More detailed analysis can be found in Appendix~\ref{AppendixC}, and the leading-order moments (therefore the leading-order intrinsic anomalous Hall responses) under all MPGs are listed in Table \ref{table:02}.  10 MPGs (out of 31) respect the $C_{2}\mathcal{T}$ symmetry, which forces the Berry curvature to vanish in the entire Brillouin zone. Among the remaining 21 MPGs, 10 of them break both the time-reversal and the mirror symmetries, therefore a non-vanishing monopole is allowed. The other 11 MPGs are also classified, according to their leading-order anomalous Hall responses.
	%	Among the remaining 21 MPGs, 10 of them  exhibit anomalous Hall effect, while 2nd-order NLAH effect becomes the leading-order Hall response in 3 MPGs. There are 3 MPGs in which the quadrupole is the leading-order moment, and the hexapole is the leading-order moment in 3 MPGs. 5th-order response becomes leading-order response in $4mm$ MPG, and 7th-order response is the leading-order response in $6mm$ MPG.
	
	\begin{table}[t]
		\begin{threeparttable}
			\caption{Leading-order intrinsic anomalous Hall responses in all the 31 2D magnetic point groups}
			\begin{ruledtabular}\label{table:02}
				\begin{tabular}{c c}
					leading-order responses & magnetic point groups \tabularnewline [0.5ex]
					\hline 
					No anomalous & $21'$, $2'$, $2mm1'$, $2'mm'$, $41'$, \tabularnewline
					Hall response & $4mm1'$, $61'$, $6'$, $6mm1'$, $6'm'm$ \tabularnewline \tabularnewline
					\multirow{2}{*}{1st-order} & $1$, $m'$, $2$, $2m'm'$, $4$, \tabularnewline
					&$4m'm'$, $3$, $3m'$, $6$, $6m'm'$ \tabularnewline \tabularnewline
					2nd-order & $11'$, $m$, $m1'$ \tabularnewline \tabularnewline
					3rd-order & $2mm$, $4'$, $4'm'm$ \tabularnewline \tabularnewline
					4th-order & $3m$, $31'$, $3m1'$ \tabularnewline \tabularnewline
					5th-order & $4mm$ \tabularnewline \tabularnewline
					7th-order & $6mm$ \tabularnewline
				\end{tabular}
			\end{ruledtabular}
		\end{threeparttable}
	\end{table}

	\begin{figure*}[t]
		\centering
		\includegraphics[width=7in]{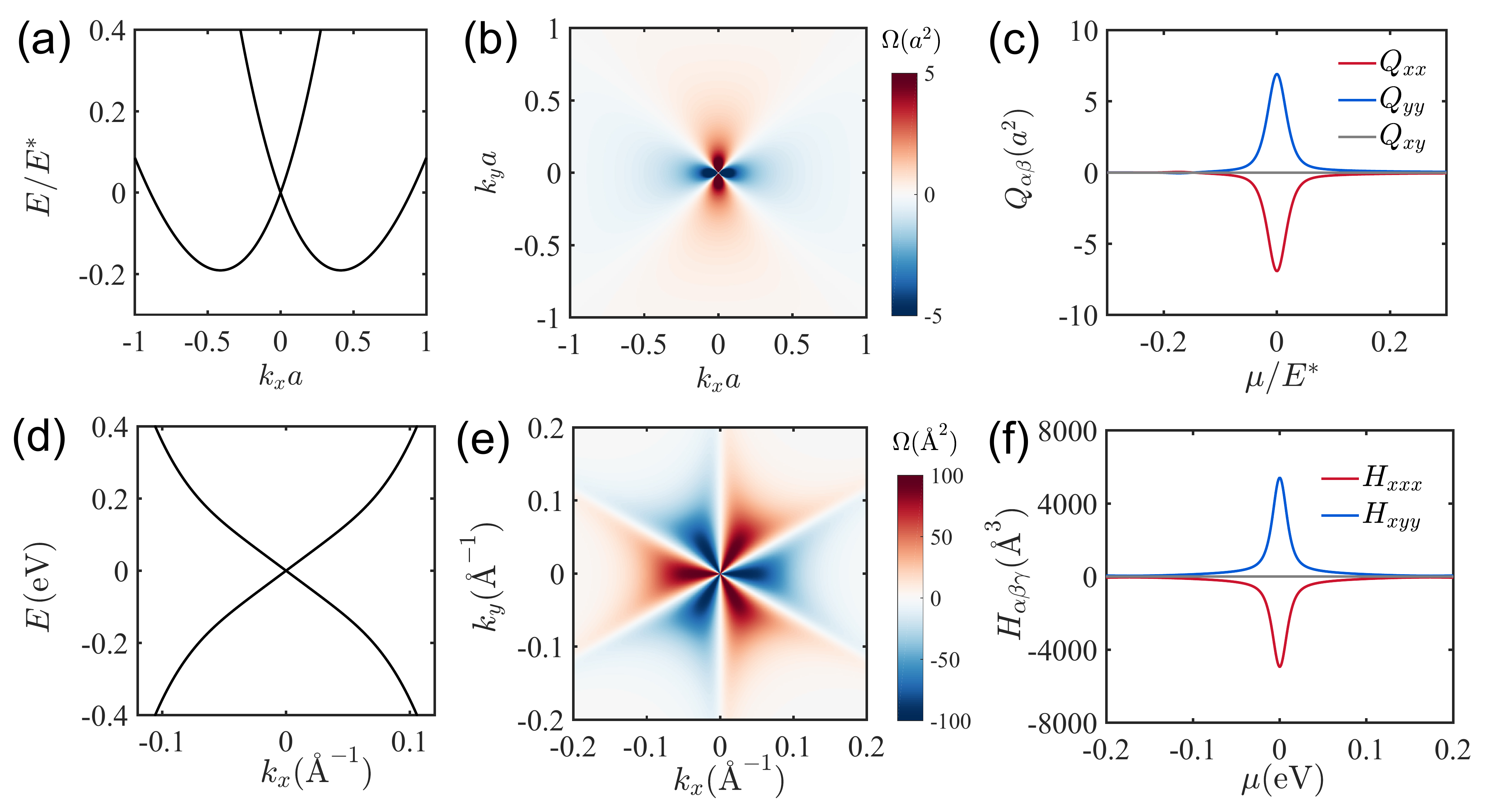}
		\caption{(a)-(c) Band structures (a),  Berry curvature of the conduction band (b) and gate dependence of the quadrupole (c) for the effective model in Eq.~\ref{eq:Hamiltonian}, with $t = 3$, $v = 1$, $m = 2$ and temperature $k_{\mathrm{B}} T = 0.005$. The length scale $a=m/v$, and the energy scale $E^{*}=v^2/m$. (d)-(f) Band structures (d),  Berry curvature of the conduction band (e) and gate dependence of the hexapole (f) for the surface states of topological insulators as described by Eq. \ref{eq:TIHamiltonian}. The parameters for Bi$_2$Te$_3$ are adopted from Ref. \cite{Fu2009hexagonal}, with $v = 2.55 \; \mathrm{eV\cdot\AA}$, $\lambda = 250 \; \mathrm{eV\cdot\AA^3}$, $E_0(k) = \alpha = 0$, and temperature $ T = 50 \; \mathrm{K}$.}
		\label{FIG01}
	\end{figure*}	
	
	\emph{Effective models.}---In this section, we use an effective model to show explicitly how the quadrupole arises as the leading-order Berry curvature moment for materials with $4'm'm$ MPG in 2D space (as indicated by symmetry analysis in Table~\ref{table:02}). Importantly, $4'm'm$ is the symmetry for antiferromagnetic monolayer SrMnBi$_2$ which will be studied in detail in the next section. Furthermore, we show that the Berry curvature hexapole can be the leading-order moment for the surface states of topological insulators~\cite{Fu2009hexagonal} with $3m1'$ ($C_{3v}$) MPG symmetry (see Table~\ref{table:02}).
		
	The $4'm'm$ MPG contains two generators: $C_{4}\mathcal{T}$ and $M_{x}\mathcal{T}$, where $C_{4}$ is the four-fold rotation around the $z$ axis and $M_{x}$ is reflection: $x \rightarrow -x$. The $C_{4}\mathcal{T}$ symmetry requires the monopole to be zero, and $C_{2}=(C_{4}\mathcal{T})^2$ also forces the dipole to vanish. In 2D space, since the Berry curvature is forced to align along the $z$ direction, the quadrupole can be denoted as $Q_{\alpha\beta} = \int_{\bm{k}}f_{0}\partial_{\alpha}\partial_{\beta}\Omega_{z}$. There are three independent components of the quadrupole: $Q_{xx}$, $Q_{yy}$, $Q_{xy}$, and their physical meanings can be understood as follows. When an AC electric current is applied along the $x(y)$ direction, $Q_{xx}$($Q_{yy}$) generates a third harmonic voltage in the $y(-x)$ direction. Furthermore, when the current is not applied along the two axes, $Q_{xy}$ will have an additional contribution to the anomalous Hall voltage. The $C_{4}\mathcal{T}$ symmetry requires $Q_{xx} = -Q_{yy}$, and the $M_{x}\mathcal{T}$ symmetry further forces $Q_{xy} = 0$. Therefore, there is only one independent non-vanishing component $Q_{xx}$ in $4'm'm$ MPG.
	
	Under $4'm'm$ MPG, we can write down an effective Hamiltonian up to the second-order in $k$ near the $\Gamma$ point:
	\begin{equation}\label{eq:Hamiltonian}
		\mathcal{H}(\bm{k}) = t k ^ 2 + v (k_{y}\sigma_{x} - k_{x}\sigma_{y}) + m (k_{x}^{2}-k_{y}^{2})\sigma_{z},
	\end{equation}
	where $\bm{\sigma}$ denotes the Pauli matrices acting on the spin degrees of freedom, and $k=|\bm{k}|$. It is a Rashba-like Hamiltonian with a second-order warping term which breaks time-reversal symmetry. The $C_{4}\mathcal{T}$ symmetry forces the bands to be doubly degenerate at the $\Gamma$ point. The energy spectra of the two bands are: $E_{\pm}(\bm{k}) = t k ^ 2 \pm |\bm{d}(\bm{k})|$, as shown in Fig. \ref{FIG01}(a). Here $\pm$ denote the conduction and valence bands respectively, and $\bm{d}(\bm{k}) = [vk_{y}, -vk_{x}, m(k_{x}^{2}-k_{y}^{2})]$.
	
	The $C_{4}\mathcal{T}$ symmetry requires $\Omega_{\pm}(\hat{C_{4}}\hat{\mathcal{T}}\bm{k}) = -\Omega_{\pm}(\bm{k})$, leading to clover-shape Berry curvature distributions as shown in Fig. \ref{FIG01}(b), which can be calculated as:
	\begin{equation}\label{eq:BC}
		\Omega_{\pm}(\bm{k}) = \pm \frac{1}{2} \hat{\bm{d}} \cdot (\partial_{x} \hat{\bm{d}} \times \partial_{y} \hat{\bm{d}}) = \mp \frac{v^2 d_{z}(\bm{k})}{2 |\bm{d}(\bm{k})| ^ 3}.
	\end{equation}
	
	Consider the situation when chemical potential $\mu$ is close to the band-crossing point: $|\mu| \ll min\{\frac{v^2}{|t|}, \frac{v^2}{|m|}\}$, where the energy dispersions are approximately linear $E_{\pm}(\bm{k}) \approx \pm vk$, and the Berry curvatures $\Omega_{\pm}(\bm{k}) \approx \mp \frac{d_{z}(\bm{k})}{2 v k ^ 3}$. The Berry curvature quadrupole at zero temperature can then be calculated with Eq. \ref{eq:BCQuadrupole2} as:
	\begin{eqnarray}
		Q_{xx} = - Q_{yy} &=& -\frac{m}{16 \pi |\mu|},\\
		Q_{xy} &=& 0.
	\end{eqnarray}
	The behavior of the Berry curvature quadrupole is depicted in Fig. \ref{FIG01}(c), which exhibits a peak near the band-crossing.
	
	It is worth noting that the quadrupoles are nearly the same for the conduction and valence bands near the band-crossing at $\mu = 0$. Since near the band-crossing point where the kinetic term $tk^2$ can be neglected, the two bands have nearly opposite energy dispersions $E_{+}(\bm{k}) \approx -E_{-}(\bm{k})$, and opposite Berry curvatures $\Omega_{+}(\bm{k}) = -\Omega_{-}(\bm{k})$. According to Eq. \ref{eq:BCQuadrupole2}, the quadrupoles of the two bands are nearly the same around the band-crossing: $Q_{\alpha\beta}^{+}(\mu) \approx Q_{\alpha\beta}^{-}(-\mu)$.

	Next, we study the Berry curvature hexapole for the surface states of topological insulators with $3m1'$ ($C_{3v}$) MPG. The time-reversal symmetry $\mathcal{T}$ requires the monopole and quadrupole to be zero, and the $C_{3}$ symmetry also forces the dipole to vanish. In 2D space,  the hexapole can be denoted as $H_{\alpha\beta\gamma} = \int_{\bm{k}}f_{0}\partial_{\alpha}\partial_{\beta}\partial_{\gamma}\Omega_{z}$. There are four independent components of the hexapole: $H_{xxx}$, $H_{xxy}$, $H_{xyy}$, $H_{yyy}$, and their physical meanings can be understood as follows. When an AC electric current is applied along the $x(y)$ direction, $H_{xxx}$($H_{yyy}$) generates a fourth harmonic voltage in the $y(-x)$ direction. Furthermore, when the current is not applied along the two axes, $H_{xxy}$ and $H_{xyy}$ will have additional contributions to the anomalous Hall voltage. The $C_{3}$ symmetry requires $H_{xxx} = -H_{xyy}$, and the $M_{x}$ symmetry further forces $H_{xxy} = H_{yyy} = 0$. Therefore, there is only one independent non-vanishing component $H_{xxx}$ in $3m1'$ ($C_{3v}$) MPG.
	
	The effective Hamiltonian up to third-order of $k$ which contains a hexagonal warping term is given by Ref. \cite{Fu2009hexagonal}:
	\begin{equation}\label{eq:TIHamiltonian}
		\mathcal{H}(\bm{k}) = E_0(k) + v_{k} (k_{x}\sigma_{y} - k_{y}\sigma_{x}) + \lambda k_{x} (k_{x}^{2}-3 k_{y}^{2})\sigma_{z},
	\end{equation}
	where $E_0(k)=\frac{k^2}{2m^{*}}$ is the kinetic energy. The velocity $v_{k}=v(1+\alpha k^2)$ could have a second-order correction, which we will neglect near the Dirac point. The energy dispersions of the two bands are: $E_{\pm}(\bm{k}) = E_0(k) \pm |\bm{d}(\bm{k})|$, as shown in Fig. \ref{FIG01}(d). Here $\bm{d}(\bm{k}) = [-vk_{y}, vk_{x}, \lambda k_{x} (k_{x}^{2}-3 k_{y}^{2})]$.
	
	The Berry curvatures of the two bands can be calculated as:
	\begin{equation}
		\Omega_{\pm}(\bm{k}) = \pm \frac{1}{2} \hat{\bm{d}} \cdot (\partial_{x} \hat{\bm{d}} \times \partial_{y} \hat{\bm{d}}) = \mp \frac{v^2 d_{z}(\bm{k})}{|\bm{d}(\bm{k})| ^ 3},
	\end{equation}
	and the Berry curvature of the conduction band is shown in Fig. \ref{FIG01}(e).
	
	When the chemical potential $\mu$ is close to the Dirac point: $|\mu| \ll  \sqrt{\frac{v^3}{\lambda}}$, the energy dispersions are approximately linear $E_{\pm}(\bm{k}) \approx \pm vk$, and the Berry curvatures $\Omega_{\pm}(\bm{k}) \approx \mp \frac{d_{z}(\bm{k})}{v k ^ 3}$. The Berry curvature hexapole at zero temperature can then be calculated as:
	\begin{eqnarray}
		H_{xxx} =& - H_{xyy} &= -\frac{3\lambda}{16 \pi |\mu|},\\
		H_{xxy} =& H_{yyy} &= 0.
	\end{eqnarray}
	The behavior of the hexapole is depicted in Fig. \ref{FIG01}(f), which has the same $|\mu|^{-1}$ dependence as the quadrupole in the antiferromagnetic model.
	
	When a current is applied along the $x$ direction which is perpendicular to the mirror plane, the hexapole will induce a NLAH voltage $V_y \propto H_{xxx}I_{x}^{4}$ along the $y$ direction. As an estimation of the NLAH voltage for Bi$_2$Te$_3$, we consider the situation when the Fermi energy is $\mu \approx 0.1 \; \mathrm{eV}$ away from the Dirac point of the surface state. Taking $\lambda = 250 \; \mathrm{eV\cdot\AA^3}$ \cite{Fu2009hexagonal}, we get the hexapole $H_{xxx} = 150 \;  \mathrm{\AA^3}$. Considering an applied electric current $\sim 20 \; \mathrm{mA}$, with the conductance $\sim 2\times10^{-3} \; \mathrm{\Omega^{-1}}$ for the surface states and the sample size $\sim 1 \; \mathrm{mm}$ \cite{Qu2010quantum}, it corresponds to an electric filed $E \sim 10 \; \mathrm{mV}/\mathrm{\mu m}$. With the scattering time $\tau \sim 0.5 \; \mathrm{ps}$ \cite{Qu2010quantum}, we obtain the NLAH current density $j^{H}(4\omega) \sim 20 \; \mathrm{pA/mm}$, which corresponds to a Hall voltage $\sim 10 \; \mathrm{nV}$.

	\emph{Candidate materials.}---In this section, we propose candidate materials to observe the third-order NLAH effect induced by Berry curvature quadrupole. Layered structure antiferromagnets AMnBi$_2$ (A = Sr, Ca, Ba, Eu) host anisotropic Dirac fermions near the Fermi surface~\cite{Park2011anisotropic, Wang2012two, Li2016electron, Masuda2016quantum}. Below the transition temperature, the MPG of bulk AMnBi$_2$ crystals is $4'/m'm'm$, which forces the bands to be doubly degenerate by the $\mathcal{I}\mathcal{T}$ symmetry. On the other hand, single-domain thin-film SrMnBi$_2$ has been fabricated on LaAlO$_3$(001) substrate~\cite{Takahashi2020single}, in which the $\mathcal{I}\mathcal{T}$ symmetry could be effectively broken by the substrate or vertical electrical gating, reducing the symmetry down to $4'm'm$ which supports nonzero quadrupole as its leading-order Berry curvature moment. To illustrate the symmetry breaking effect, here we study the Berry curvature quadrupole in monolayer (one sextuple layer) SrMnBi$_2$ with first-principles calculations.
	
\begin{figure}
		\centering
		\includegraphics[width=3.5in]{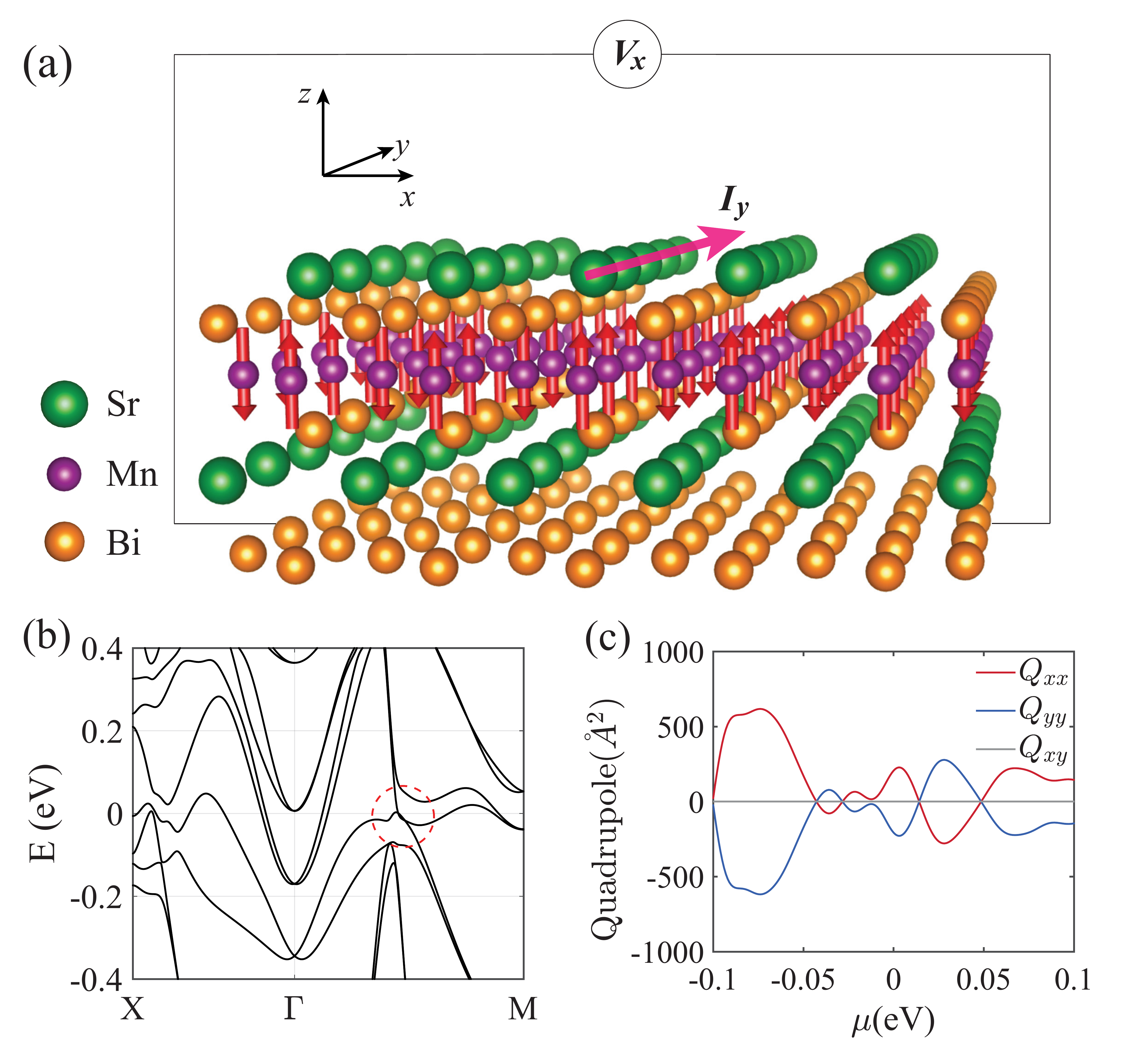}
		\caption{(a) Crystal structure of monolayer (one sextuple layer) SrMnBi$_2$, with the Mn atoms exhibit antiferromagnetic order as indicated by the red arrows. When a current is applied along the $y$ direction, a NLAH voltage $V_x$ will be induced by the Berry curvature quadrupole. (b) Band structures of monolayer SrMnBi$_2$. Red dashed circle indicates the Dirac cone. (c) Berry curvature quadrupole of SrMnBi$_2$ near the Fermi level, with temperature $T = 100 \; \mathrm{K}$.}
		\label{FIG02}
	\end{figure}

	The crystal structure of monolayer SrMnBi$_2$ is shown in Fig.~\ref{FIG02}(a). It contains a Mn layer which exhibits the antiferromagnetic order, and a conducting Bi layer which provides the Dirac fermions at the Fermi level~\cite{Park2011anisotropic}. The Dirac fermions are located along the $\Gamma-\mathrm{M}$ lines, as indicated by the red dashed circle in Fig.~\ref{FIG02}(b). The $\mathcal{I}\mathcal{T}$ symmetry coming from interlayer stacking in bulk crystals is absent in the monolayer, which lifts the two-fold degeneracy of the bands. Moreover, a small gap can be opened by the spin-orbit coupling~\cite{Park2011anisotropic}, which generates Berry curvatures near the band edges. When the chemical potential is near the Dirac point, sizable Berry curvature quadrupole $\sim 500 \; \mathrm{\AA^2}$ can be obtained, as shown in Fig.~\ref{FIG02}(c).
	
	%	The Dirac fermions in SrMnBi$_2$ are highly anisotropic, with the Fermi velocity along the $\Gamma-\mathrm{M}$ line to be $\sim 8$ times larger than the Fermi velocity along the perpendicular direction \cite{Park2011anisotropic}. As a result, the bands are always nearly touching along the perpendicular direction, leading to sizable Berry curvature all the way to the $\Gamma-\mathrm{X}$ line.
	
	When an electric current is applied along the $y$ direction, the quadrupole will induce a NLAH voltage $V_x \propto Q_{yy}I_{y}^{3}$ along the $x$ direction, as shown in Fig.~\ref{FIG02}(a). As an estimation of the Hall voltage, we take an experimentally accessible current $\sim 100 \; \mathrm{\mu A}$~\cite{Ma2019observation, Kang2019nonlinear}. With the resistance $\sim 1 \; \mathrm{k\Omega}$~\cite{Resistance}, sample size $\sim 10 \; \mathrm{\mu m}$, it corresponds to an applied electric filed $E \sim 10 \; \mathrm{mV}/\mathrm{\mu m}$.  By taking the quadrupole $\sim 500 \; \mathrm{\AA^2}$ and the scattering time $\tau \sim 1 \; \mathrm{ps}$, we obtain the NLAH current density $j^{H}(3\omega) \sim 1 \; \mathrm{nA/\mu m}$, corresponding to a Hall voltage $\sim 10 \; \mathrm{\mu V}$, which is comparable with the nonlinear Hall voltage in WTe$_2$~\cite{Ma2019observation, Kang2019nonlinear}.
		
	In ferromagnets, although the Berry curvature quadrupole is not the leading-order moment, it can still have contributions to the NLAH voltage. As the quadrupole contributes to the third harmonic generation, it could be distinguished from the anomalous Hall response with lock-in techniques. 2D magnets MnBi$_2$Te$_4$~\cite{Deng2020quantum}, Fe$_3$GeTe$_2$~\cite{Deng2018gate}, near 3/4-filling twisted bilayer graphene \cite{Sharpe2019emergent, Serlin2020intrinsic} and 3D antiferromagnetic Weyl semimetals~\cite{Weyl} Mn$_3$X (X = Ge, Sn, Ga, Ir, Rh, Pt)~\cite{Chen2014anomalous, Kubler2014non-collinear, Nakatsuji2015large, Nayak2016large, Zhang2017strong}, GdPtBi \cite{Suzuki2016large} all have non-vanishing quadrupoles. Especially, MnBi$_2$Te$_4$ and twisted bilayer graphene exhibit quantum anomalous Hall effect~\cite{Deng2020quantum, Fu2020exchange, Sharpe2019emergent, Serlin2020intrinsic}, which are therefore good platforms to study the Berry curvature effects. Sizable Berry curvature quadrupoles $\sim 4000 \; \mathrm{\AA^{2}}$ are found for twisted bilayer graphene near its ferromagnetic quantum anomalous Hall phase, as discussed in details in Appendix~\ref{AppendixD}.
	
	%Furthermore, heterostructures formed by magnetic materials such as CrI$_3$ with WTe$_2$ ~\cite{Zhong2017van, Zhao2020magnetic} are also promising platforms to observe the NLAH effect. The magnetic layer breaks time-reversal symmetry, and at the same time the topological material can provide nontrivial Berry curvature near the Fermi surface. Due to the flexibility of engineering heterostructures, it provides abundant possibilities to observe the NLAH effect.

	\emph{Discussions and conclusions.}---In this work, we establish a general theory for the higher-order NLAH effects induced by Berry curvature multipoles. In particular, we point out that the third-order NLAH effect can be the leading-order Hall response in certain antiferromagnets, and the fourth-order NLAH can be the leading response in the surface states of topological insulators. We also propose candidate materials including 2D antiferromagnets and ferromagnets, Weyl semimetals and twisted bilayer graphene near the quantum anomalous Hall phase to observe the third-order NLAH effect.
	
	Here, we further discuss several issues about the higher-order NLAH effects induced by Berry curvature multipoles. First of all, we discuss how to distinguish the NLAH responses related to the anti-symmetric part of the conductivity tensor from the Drude-like contributions which are related to the symmetric part of the conductivity tensor. For even-order responses, the Drude-like contribution is forbidden by the time-reversal symmetry, which is the case for the surface states of topological insulators. For odd-order responses, if a mirror symmetry $M_{x}$ or $M_{x}\mathcal{T}$ is present, the Drude-like contribution vanishes in the transverse direction when the current is applied perpendicular to the mirror plane \cite{Parker2019diagrammatic}. In monolayer SrMnBi$_2$ which belongs to $4'm'm$ MPG, the third-order Drude-like contribution vanishes when the current is applied along the $x$ or the $y$ directions, as indicated in Fig.~\ref{FIG02}(a). In general cases, the NLAH responses can be distinguished from the Drude-like contributions as they have different angular dependence, and the details can be found in Appendix~\ref{AppendixE}.
	
	Second, we note that apart from the intrinsic contributions from Berry curvature, impurities can also give rise to the NLAH voltage through skew scattering and side jump~\cite{Nagaosa2010anomalous, Nandy2019symmetry, Du2019disorder, Xiao2019theory}. Importantly, these contributions have the same symmetry properties~\cite{Nagaosa2010anomalous, Nandy2019symmetry}, thus will not affect the symmetry discussions in previous sections.
	
	Third, besides the nonlinear Hall effect, it is shown that the Berry curvature dipole can induce other nonlinear effects such as the nonlinear Nernst~\cite{yu2019topological, zeng2019nonlinear} and nonlinear thermal Hall effect~\cite{zeng2020fundamental}. Here, we believe that the Berry curvature multipoles can also contribute to the corresponding higher-order nonlinear Nernst and nonlinear thermal Hall effects.
	
	\emph{Acknowledgments.}---The authors thank Benjamin T. Zhou and Mengli Hu for valuable discussions. KTL acknowledges the support of the Croucher Foundation, the Dr. Tai-chin Lo Foundation and the HKRGC through grants C6025-19G, 16310219 and 16309718. HCP acknowledges support from a Papparlardo Fellowship at MIT.
	
	\emph{Note added.}---Recently, it was pointed out by He et al.~\cite{He2020quantum} that the surface states of topological insulators can be used for second harmonic generation in which an applied AC current can induce a voltage with double frequency. This effect is caused by disorder induced skew scattering and related to the symmetric part of the nonlinear conductivity tensor. On the other hand, the fourth-order NLAH effect on the surface states of topological insulator discussed in this work is an intrinsic effect induced by Berry curvature hexapole which is related to the anti-symmetric part of the nonlinear conductivity tensor. These two effects have different physical origins and frequency dependence and they can be distinguished experimentally.
	
	\appendix
	\renewcommand{\theequation}{A-\arabic{equation}}
	\renewcommand\thefigure{A-\arabic{figure}} 
	\setcounter{equation}{0}  
	\setcounter{figure}{0}
	\section{First-principles calculations}\label{AppendixA}
	
	In this work, the density functional theory (DFT) computations were performed by utilizing the \textit{Vienna Ab initio Simulation Package} (VASP) \cite{kresse1996efficiency} with the projector-augmented wave method \cite{blochl1994projector} and the Perdew-Berke-Ernzerhof’s (PBE) exchange-correlation functional in the generalized-gradient approximation (GGA) \cite{perdew1996generalized}. Specifically for the two antiferromagnetic materials CaMnBi$_2$ and SrMnBi$_2$, the LDA+U approach \cite{anisimov1991band} was adopted to modify the intra-atomic Coulomb interaction which is essential for magnetism. We used an empirical value $U_{eff}=3$ eV for the $d$-orbitals of Mn atoms. For computing the electronic bands of the monolayer materials, a vacuum layer of thickness 20 \AA \ was added along the $z$-direction. A $9\times9\times9$ $k$ mesh grid was used in the self-consistent calculation step for the bulk material, while for the monolayer cases we adopted a $9\times9\times1$ $k$ mesh grid.
	
	In the calculation of the Berry curvature and the Berry curvature quadrupoles, we adopted two methods and the results of the two methods were consistent with each other. In the first method, we adapted the VASPBERRY codes which can compute the Berry curvature and the Chern numbers in 2D systems directly using the VASP wavefunctions via Fukui's method \cite{fukui2005chern}. For the second approach, the maximally localized Wannier bands of CaMnBi$_2$ and SrMnBi$_2$ were projected through the Wannier90 package \cite{mostofi2014updated} linked to VASP, based on which the Berry curvature was computed.
	
\begin{figure}
		\centering
		\includegraphics[width=3.5in]{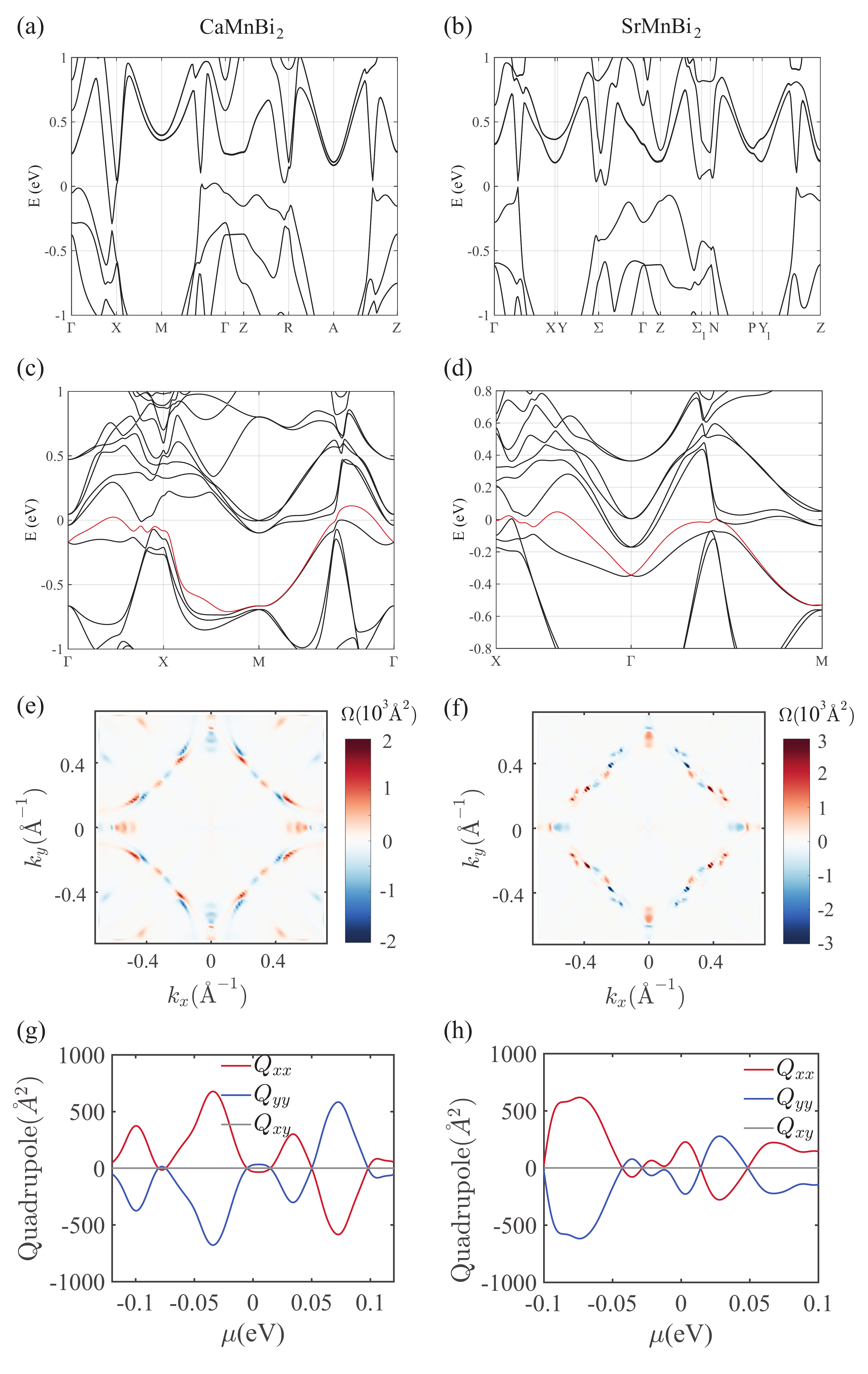}
		\caption{(a) \& (b) The band structures of bulk CaMnBi$_2$ and SrMnBi$_2$ in the antiferromagnetic phase based on DFT calculations. (c) \& (d) The band structures of monolayer (one sextuple layer) CaMnBi$_2$ and SrMnBi$_2$ in the antiferromagnetic phase. (e) \& (f) The Berry curvature configuration in the 2D BZ of monolayer CaMnBi$_2$ and SrMnBi$_2$ for the top valence bands, which are denoted as red curves in (c) \& (d). (g) \& (h) The Berry curvature quadrupoles of monolayer CaMnBi$_2$ and SrMnBi$_2$ as functions of the chemical potential.}
		\label{FIG_DFT}
	\end{figure}

	To be concise, we have only shown the results of the calculations for monolayer SrMnBi$_2$. In this section, we provide more related DFT results, including the band structures of bulk and monolayer Ca(Sr)MnBi$_2$, the Berry curvature configuration in the 2D Brillouin zone (BZ) for the bands near the Fermi energy in monolayer Ca(Sr)MnBi$_2$, as well as the Berry curvature quadrupole values at different chemical potentials, as shown in Fig.~\ref{FIG_DFT}.
	
	Notably, due to the anisotropy of the Dirac cones~\cite{Park2011anisotropic, S_Feng2014strong}, i.e., the Fermi velocity along the direction perpendicular to the $\Gamma-\mathrm{M}$ line is much smaller than the Fermi velocity along the $\Gamma-\mathrm{M}$ line, the bands are nearly touching along the perpendicular direction. As a result, sizable Berry curvature is always present along the perpendicular direction, as shown in Fig.~\ref{FIG_DFT} (e) and (f), leading to the large Berry curvature quadrupole near the Fermi energy.
	
	\renewcommand{\theequation}{B-\arabic{equation}}
	\renewcommand\thefigure{B-\arabic{figure}} 
	\setcounter{equation}{0}  
	\setcounter{figure}{0}
	\section{Nonlinear conductivity and Berry curvature multipoles}\label{AppendixB}
	
	In this appendix, we establish the relationship between the nonlinear conductivity in the presence of an AC electric field and the Berry curvature multipoles. We focus on the intra-band contributions, and this approximation  is valid when the frequency is much lower than the energy gap of the bands near the Fermi energy. A single band is considered in the derivation, which can be generalized to multi-band cases by summing up the contributions from all the bands.
	
	First of all, the electric current density is related the integral of physical velocities of the electrons:
	\begin{equation}\label{eq:current}
		\bm{j}(t)  = -e \int_{\bm{k}} f(\bm{k},t) \frac{d\bm{r}}{dt},	
	\end{equation}
	where $\int_{\bm{k}} = \int d^d k/(2\pi)^d$, and $d$ is the dimensionality. The physical velocity of an electron has two contributions, which are the group velocity of the electron wave packet and the anomalous velocity originating from the Berry curvature:
	\begin{equation}
		\frac{d}{dt}\bm{r} = \frac{1}{\hbar} \nabla_{\bm{k}}\varepsilon_{\bm{k}} + \frac{e}{\hbar} \bm{E} \times \bm{\Omega},
	\end{equation}
	where $\varepsilon_{\bm{k}}$ is the energy dispersion, and $\bm{E}=\bm{E}(t)$ is the applied time-dependent electric field, $\bm{\Omega}$ is the vector form of Berry curvature, which is related to its tensor representation $\Omega_{\alpha}=\frac{1}{2}\epsilon_{\alpha\beta\gamma}\mathcal{F}_{\beta\gamma}$, and
	\begin{equation}
		\mathcal{F}_{\alpha\beta}=\partial_{\alpha}\mathcal{A}_{\beta}-\partial_{\beta}\mathcal{A}_{\alpha}, \quad \mathcal{A}_{\alpha}= -i \bra{u_{\bm{k}}} \partial_{\alpha} \ket{u_{\bm{k}}}.
	\end{equation}
	Here, $\ket{u_{\bm{k}}}$ is the periodic part of the Bloch wave function at $\bm{k}$, and $\partial_{\alpha} = \partial/\partial_{k_{\alpha}}$. $\epsilon_{\mu\alpha\beta}$ is the Levi-Civita tensor, and the Greek letters $\alpha, \beta, \gamma = x, y, z$ represent the spatial indices.
	%	
	%	Substituting the physical velocity into Eq.~\ref{eq:current}, we obtain the current response, which contains an anomalous Hall response
	%	\begin{equation}\label{Hall}
	%		j_{\mu}^{H}(t)=-\frac{e^2}{\hbar}\int_{\bm{k}}\mathcal{F}_{\mu\alpha}E_{\alpha}(t)f(\boldsymbol{k},t),
	%	\end{equation}
	%	and a Drude-like contribution
	%	\begin{equation}\label{Drude}
	%		j_{\mu}^{D}(t)=-\frac{e}{\hbar}\int_{\bm{k}} \partial_{\mu}\varepsilon_{\bm{k}}f(\boldsymbol{k},t).
	%	\end{equation}
	
	The time evolution of the distribution function $f(\bm{k},t)$ is given by the Boltzmann equation in the relaxation time approximation:
	\begin{equation}\label{eq:Boltzmann}
		\frac{d\bm{k}}{dt} \cdot \nabla_{\bm{k}} f(\bm{k},t) + \partial_{t} f(\bm{k},t) = \frac{f_{0}-f(\bm{k},t)}{\tau},
	\end{equation}
	where $f_0$ is the equilibrium Fermi-Dirac distribution function and $\tau$ represents the relaxation time. The evolution of the canonical momentum is given by the semiclassical equation of motion:
	\begin{equation}
		\frac{d}{dt}\bm{k} = -\frac{e\bm{E}}{\hbar}.
	\end{equation}
	
	With a time-dependent electric field $\bm{E}(t)=\mathrm{Re}\{E_{\alpha}e^{i\omega t}\bm{\hat{e}}_{\alpha}\}$, we calculate the current response order by order. The
	distribution function can be expanded in orders of the electric field:
	\begin{equation}
		f(\bm{k},t)=\mathrm{Re}\{\sum_{n=0}^{\infty} f_n(\bm{k},t)\},
	\end{equation}
	where the term $f_n$ is proportional to $E^n$. Substituting into Eq.~\ref{eq:Boltzmann}, we obtain a recursive equation for the adjacent orders of $f_n$:
	\begin{equation}\label{eq:recurcive}
		(\partial_{t}+1/\tau)\mathrm{Re}\{f_{n+1}\}=\frac{e}{\hbar}\mathrm{Re}\{E_{\alpha}e^{i\omega t}\}\mathrm{Re}\{\partial_{\alpha}f_{n}\}.
	\end{equation}
	$f_n$ can be further decomposed according to the frequency dependence:
	\begin{equation}
		f_{n} =\sum_{m=0}^{\infty} f_{n}(m\omega)e^{im\omega t}.
	\end{equation}
	Starting from the zeroth-order, all the $f_{n}(m\omega)$ components can be calculated recursively. The first and second-order nonzero terms are
	\begin{eqnarray}
		f_{1}(\omega) & = & \frac{e}{\hbar}\frac{\partial_{\alpha}f_{0}}{\widetilde{\omega}}E_{\alpha},\\\nonumber
		f_{2}(0) & = & \left(\frac{e}{\hbar}\right)^{2}\frac{\partial_{\alpha}\partial_{\beta}f_{0}}{2\gamma(\widetilde{2\omega})}E_{\alpha}E_{\beta}^{*},\\
		f_{2}(2\omega)  & = & \left(\frac{e}{\hbar}\right)^{2}\frac{\partial_{\alpha}\partial_{\beta}f_{0}}{2\widetilde{\omega}(\widetilde{2\omega})}E_{\alpha}E_{\beta},
	\end{eqnarray}
	as first obtained by Sodemann and Fu~\cite{Sodemann2015quantum}. Here, $\widetilde{m\omega}$ represents $im\omega+\gamma$ and $\gamma=1/\tau$. The third-order terms are
	\begin{eqnarray}\nonumber
		f_{3}(\omega) & = & \left(\frac{e}{\hbar}\right)^{3}\frac{3\partial_{\alpha}\partial_{\beta}\partial_{\gamma}f_{0}}{4\widetilde{\omega}(\widetilde{-\omega})(\widetilde{2\omega})}E_{\alpha}E_{\beta}E_{\gamma}^{*},\\
		f_{3}(3\omega) & = & \left(\frac{e}{\hbar}\right)^{3}\frac{\partial_{\alpha}\partial_{\beta}\partial_{\gamma}f_{0}}{4\widetilde{\omega}(\widetilde{2\omega})(\widetilde{3\omega})}E_{\alpha}E_{\beta}E_{\gamma}.
	\end{eqnarray}
	Substituting the distribution function into Eq.~\ref{eq:current}, we obtain the current response  $\bm{j}(t)=\sum\limits_{n=1}^{\infty} \bm{j}^{(n)}(t)$ in orders of the electric field. The linear response is at the same frequency of the driving force: 
	\begin{equation}
		j^{(1)}_{\mu}(t)=\mathrm{Re}\{\sigma^{(1)}_{\mu\alpha}(\omega)E_{\alpha}e^{i\omega t}\}, 
	\end{equation}
	with
	\begin{equation}
		\sigma^{(1)}_{\mu\alpha}(\omega)=\frac{e^{2}}{\hbar^{2}}\int_{\bm{k}}f_{0}\frac{\partial_{\mu}\partial_{\alpha}\varepsilon_{\bm{k}}}{\widetilde{\omega}}-\frac{e^{2}}{\hbar}\int_{\bm{k}}f_{0}\mathcal{F}_{\mu\alpha}.
	\end{equation}
	The first term is the usual Drude conductivity, which is symmetric with respect to the two indices: $\sigma^{(1), D}_{\mu\alpha}(\omega) = \sigma^{(1), D}_{\alpha\mu}(\omega)$. The second term is the intrinsic contribution to the anomalous Hall conductivity induced by Berry curvature monopole, which is anti-symmetric with respect to the two indices: $\sigma^{(1), H}_{\mu\alpha}(\omega) = - \sigma^{(1), H}_{\alpha\mu}(\omega)$.
	
	The second-order response consists of a rectified current and a second harmonic generation: 
	\begin{equation}
		j^{(2)}_{\mu}(t)=\mathrm{Re}\{\sigma^{(2)}_{\mu\alpha\beta}(0)E_{\alpha}E_{\beta}^{*}+\sigma^{(2)}_{\mu\alpha\beta}(2\omega)E_{\alpha}E_{\beta}e^{2i\omega t}\},
	\end{equation}
	with
	\begin{eqnarray}\nonumber
		\sigma^{(2)}_{\mu\alpha\beta}(0)&=&-\frac{e^{3}}{2\hbar^{3}}\int_{\bm{k}}f_{0}\frac{\partial_{\mu}\partial_{\alpha}\partial_{\beta}\varepsilon_{\bm{k}}}{\gamma\widetilde{\omega}}\\
		&&+\frac{e^{3}}{2\hbar^{2}}\int_{\bm{k}}f_{0}\frac{\partial_{\beta}\mathcal{F}_{\mu\alpha}}{\widetilde{\omega}},\\\nonumber
		\sigma^{(2)}_{\mu\alpha\beta}(2\omega)&=&-\frac{e^{3}}{2\hbar^{3}}\int_{\bm{k}}f_{0}\frac{\partial_{\mu}\partial_{\alpha}\partial_{\beta}\varepsilon_{\bm{k}}}{\widetilde{\omega}(\widetilde{2\omega})}\\
		&&+\frac{e^{3}}{2\hbar^{2}}\int_{\bm{k}}f_{0}\frac{\partial_{\beta}\mathcal{F}_{\mu\alpha}}{\widetilde{\omega}}.
	\end{eqnarray}
	Each conductivity tensor contains two terms. The first term is the Drude-like contribution and the second term is the nonlinear Hall conductivity induced by Berry curvature dipole~\cite{Sodemann2015quantum}.
	
	The third-order response is composed of currents at both the same and triple the fundamental frequency: 
	\begin{eqnarray}\nonumber
		j^{(3)}_{\mu}(t)&=&\mathrm{Re}\{\sigma^{(3)}_{\mu\alpha\beta\gamma}(\omega)E_{\alpha}E_{\beta}E_{\gamma}^{*}e^{i\omega t}\\
		&&+\sigma^{(3)}_{\mu\alpha\beta\gamma}(3\omega)E_{\alpha}E_{\beta}E_{\gamma}e^{3i\omega t}\},
	\end{eqnarray}
	with
	\begin{eqnarray}\nonumber
		\sigma^{(3)}_{\mu\alpha\beta\gamma}(\omega)&=&\frac{3e^{4}}{4\hbar^{4}}\int_{\bm{k}}f_{0}\frac{\partial_{\mu}\partial_{\alpha}\partial_{\beta}\partial_{\gamma}\varepsilon_{\bm{k}}}{\widetilde{\omega}(\widetilde{-\omega})(\widetilde{2\omega})}\\\nonumber
		&&-\frac{e^{4}}{4\hbar^{3}}\int_{\bm{k}}f_{0}[\frac{2\partial_{\beta}\partial_{\gamma}\mathcal{F}_{\mu\alpha}}{\widetilde{\omega}(\widetilde{-\omega})}+\frac{\partial_{\alpha}\partial_{\beta}\mathcal{F}_{\mu\gamma}}{\widetilde{\omega}(\widetilde{2\omega})}],\\
		&&\\\nonumber
		\sigma^{(3)}_{\mu\alpha\beta\gamma}(3\omega)&=&\frac{e^{4}}{4\hbar^{4}}\int_{\bm{k}}f_{0}\frac{\partial_{\mu}\partial_{\alpha}\partial_{\beta}\partial_{\gamma}\varepsilon_{\bm{k}}}{\widetilde{\omega}(\widetilde{2\omega})(\widetilde{3\omega})}\\
		&&-\frac{e^{4}}{4\hbar^{3}}\int_{\bm{k}}f_{0}\frac{\partial_{\beta}\partial_{\gamma}\mathcal{F}_{\mu\alpha}}{\widetilde{\omega}(\widetilde{2\omega})}.
	\end{eqnarray}
	The first term is the Drude-like contribution and the second term is the NLAH conductivity induced by Berry curvature quadrupole~\cite{Parker2019diagrammatic}.
	
	In general, the $n$-th order response has components at frequency $n\omega$, $(n-2)\omega$, $\cdots$. For the higher-order effects, we only focus on the response at frequency $n\omega$. From the recursive Eq.~\ref{eq:recurcive}, we obtain
	\begin{equation}
		f_{n}(n\omega)=2\left(\frac{e}{2\hbar}\right)^{n}\prod_{m=1}^{n}\left(E_{\alpha_{m}}\frac{\partial_{\alpha_{m}}}{\widetilde{m\omega}}\right)f_{0}.
	\end{equation}
	Substituting the distribution function into Eq.~\ref{eq:current}, we obtain the $n$-th harmonic generation
	\begin{equation}
		j^{(n)}_{\mu}(n\omega t)=\mathrm{Re}\{\sigma^{(n)}_{\mu\alpha_1\alpha_2\cdots\alpha_n}(n\omega) E_{\alpha_1} E_{\alpha_2} \cdots E_{\alpha_n} e^{in\omega t}\},
	\end{equation}
	with
	\begin{eqnarray}\nonumber
		\sigma^{(n)}_{\mu\alpha_1\alpha_2\cdots\alpha_n}(n\omega)&=&\frac{4\left(-\frac{e}{2\hbar}\right)^{n+1}}{\prod\limits_{m=1}^{n}\widetilde{m\omega}}\int_{\bm{k}}f_{0}\partial_{\mu}\partial_{\alpha_1}\partial_{\alpha_2} \cdots \partial_{\alpha_n}\varepsilon_{\bm{k}}\\\nonumber
		&&+\frac{2e\left(-\frac{e}{2\hbar}\right)^{n}}{\prod\limits_{m=1}^{n-1}\widetilde{m\omega}}\int_{\bm{k}}f_{0}\partial_{\alpha_2} \cdots \partial_{\alpha_n}\mathcal{F}_{\mu\alpha_1}.\\
		&&
	\end{eqnarray}
	By defining the ($n-1$)-th Berry curvature moment
	\begin{equation}
		P_{\alpha_{2}\cdots\alpha_{n}\beta}=\int_{\bm{k}}f_{0}\partial_{\alpha_{2}}\cdots\partial_{\alpha_{n}}\Omega_{\beta},
	\end{equation}
	we can establish the relation between the NLAH conductivity tensor and the Berry curvature moments
	\begin{equation}
		\sigma^{(n),H}_{\mu\alpha_1\alpha_2\cdots\alpha_n}(n\omega)=\frac{2e\left(-\frac{e}{2\hbar}\right)^{n}}{\prod\limits_{m=1}^{n-1}\widetilde{m\omega}}\epsilon_{\mu\alpha_{1}\beta}P_{\alpha_{2}\cdots\alpha_{n}\beta}.
	\end{equation}
	
	\renewcommand{\theequation}{C-\arabic{equation}}
	\renewcommand\thefigure{C-\arabic{figure}} 
	\setcounter{equation}{0}  
	\setcounter{figure}{0}
	\section{Symmetry analysis for Berry curvature multipoles}\label{AppendixC}
	
	\subsection{Explicit forms of Berry curvature quadrupoles under magnetic point groups}
	As shown in previous sections, the Berry curvature quadrupoles have the same transformation properties as the piezomagnetic tensors, whose explicit forms are determined under all the MPGs. Here, we list the general forms in Table \ref{table:S01}, which is quoted from Ref. \cite{Newnham2005properties}. The 18 components of the Berry curvature quadrupoles are organized as follows: 
	$\begin{pmatrix}
		Q_{xxx} & Q_{yyx} & Q_{zzx} & Q_{yzx} & Q_{xzx} & Q_{xyx}\\
		Q_{xxy} & Q_{yyy} & Q_{zzy} & Q_{yzy} & Q_{xzy} & Q_{xyy}\\
		Q_{xxz} & Q_{yyz} & Q_{zzz} & Q_{yzz} & Q_{xzz} & Q_{xyz}
	\end{pmatrix}$.

	We wish to point out that, under $4'm'm$ MPG, we have shown in previous sections that the symmetries require $Q_{xxz} = -Q_{yyz}$ and $Q_{xyz} = 0$, which is related to the result in this table by a $45 \degree$ rotation.

	\begingroup
	\setlength{\LTcapwidth}{\textwidth}
	\setlength\tabcolsep{15pt}
	\begin{longtable*}{c c}
		\\ [5ex]
		\caption{Berry curvature quadrupoles for all the magnetic point groups} \label{table:S01} \\
		
		\hline\hline \\ [0.5ex] \multicolumn{1}{c}{magnetic point groups} & \multicolumn{1}{c}{forms of Berry curvature quadrupoles} \\ [0.5ex] \hline \\ [0.5ex]
		\endfirsthead
		
		\multicolumn{2}{c}%
		{{ \tablename\ \thetable{} -- Continued from previous page}} \\ [0.5ex]
		\hline\hline \\ [0.5ex] \multicolumn{1}{c}{magnetic point groups} & \multicolumn{1}{c}{forms of Berry curvature quadrupoles} \\ [0.5ex] \hline \\ [0.5ex]
		\endhead
		
		\\ [0.5ex] \multicolumn{2}{r}{{Continued on next page}} \\ [0.5ex] \hline\hline
		\endfoot
		
		\hline \hline
		\endlastfoot
		
		$1$, $\bar{1}$ & 
		$\begin{pmatrix}
			Q_{11} & Q_{12} & Q_{13} & Q_{14} & Q_{15} & Q_{16}\\
			Q_{21} & Q_{22} & Q_{23} & Q_{24} & Q_{25} & Q_{26}\\
			Q_{31} & Q_{32} & Q_{33} & Q_{34} & Q_{35} & Q_{36}
		\end{pmatrix}$\tabularnewline\tabularnewline [1ex]
		$2$, $m$, $2/m$ & 
		$\begin{pmatrix}
			0 & 0 & 0 & Q_{14} & 0 & Q_{16}\\
			Q_{21} & Q_{22} & Q_{23} & 0 & Q_{25} & 0\\
			0 & 0 & 0 & Q_{34} & 0 & Q_{36}
		\end{pmatrix}$\tabularnewline\tabularnewline [1ex]
		$2'$, $m'$, $2'/m'$ & 
		$\begin{pmatrix}
			Q_{11} & Q_{12} & Q_{13} & 0 & Q_{15} & 0\\
			0 & 0 & 0 & Q_{24} & 0 & Q_{26}\\
			Q_{31} & Q_{32} & Q_{33} & 0 & Q_{35} & 0
		\end{pmatrix}$\tabularnewline\tabularnewline [1ex]
		$222$, $mm2$, $mmm$ & 
		$\begin{pmatrix}
			0 & 0 & 0 & Q_{14} & 0 & 0\\
			0 & 0 & 0 & 0 & Q_{25} & 0\\
			0 & 0 & 0 & 0 & 0 & Q_{36}
		\end{pmatrix}$\tabularnewline\tabularnewline [1ex]
		$2'2'2$, $m'm'2$, $m'2'm$, $m'm'm$ & 
		$\begin{pmatrix}
			0 & 0 & 0 & 0 & Q_{15} & 0\\
			0 & 0 & 0 & Q_{24} & 0 & 0\\
			Q_{31} & Q_{32} & Q_{33} & 0 & 0 & 0
		\end{pmatrix}$\tabularnewline\tabularnewline [1ex]
		$3$, $\bar{3}$ & 
		$\begin{pmatrix}
			Q_{11} & -Q_{11} & 0 & Q_{14} & Q_{15} & -2Q_{22}\\
			-Q_{22} & Q_{22} & 0 & Q_{15} & -Q_{14} & -2Q_{11}\\
			Q_{31} & Q_{31} & Q_{33} & 0 & 0 & 0
		\end{pmatrix}$\tabularnewline\tabularnewline [1ex]
		$32$, $3m$, $\bar{3}m$ & 
		$\begin{pmatrix}
			Q_{11} & -Q_{11} & 0 & Q_{14} & 0 & 0\\
			0 & 0 & 0 & 0 & -Q_{14} & -2Q_{11}\\
			0 & 0 & 0 & 0 & 0 & 0
		\end{pmatrix}$\tabularnewline\tabularnewline [1ex]
		$32'$, $3m'$, $\bar{3}m'$ & 
		$\begin{pmatrix}
			0 & 0 & 0 & 0 & Q_{15} & -2Q_{22}\\
			-Q_{22} & Q_{22} & 0 & Q_{15} & 0 & 0\\
			Q_{31} & Q_{31} & Q_{33} & 0 & 0 & 0
		\end{pmatrix}$\tabularnewline\tabularnewline [1ex]
		$4$, $\bar{4}$, $4/m$, $6$, $\bar{6}$, $6/m$ & 
		$\begin{pmatrix}
			0 & 0 & 0 & Q_{14} & Q_{15} & 0\\
			0 & 0 & 0 & Q_{15} & -Q_{14} & 0\\
			Q_{31} & Q_{31} & Q_{33} & 0 & 0 & 0
		\end{pmatrix}$\tabularnewline\tabularnewline [1ex]
		$4'$, $\bar{4}'$, $4'/m$ & 
		$\begin{pmatrix}
			0 & 0 & 0 & Q_{14} & Q_{15} & 0\\
			0 & 0 & 0 & -Q_{15} & Q_{14} & 0\\
			Q_{31} & -Q_{31} & 0 & 0 & 0 & Q_{36}
		\end{pmatrix}$\tabularnewline\tabularnewline [1ex]
		$422$, $4mm$, $\bar{4}2m$, $4/mmm$, $622$, $6mm$, $\bar{6}m2$, $6/mmm$ & 
		$\begin{pmatrix}
			0 & 0 & 0 & Q_{14} & 0 & 0\\
			0 & 0 & 0 & 0 & -Q_{14} & 0\\
			0 & 0 & 0 & 0 & 0 & 0
		\end{pmatrix}$\tabularnewline\tabularnewline [1ex]
		$4'22$, $4'm'm$, $\bar{4}'2m'$, $\bar{4}'2'm$, $4'/mmm'$ & 
		$\begin{pmatrix}
			0 & 0 & 0 & Q_{14} & 0 & 0\\
			0 & 0 & 0 & 0 & Q_{14} & 0\\
			0 & 0 & 0 & 0 & 0 & Q_{36}
		\end{pmatrix}$\tabularnewline\tabularnewline [1ex]
		$42'2'$, $4m'm'$, $\bar{4}2'm'$, $4/mm'm'$, $62'2'$, $6m'm'$, $\bar{6}m'2'$, $6/mm'm'$ & 
		$\begin{pmatrix}
			0 & 0 & 0 & 0 & Q_{15} & 0\\
			0 & 0 & 0 & Q_{15} & 0 & 0\\
			Q_{31} & Q_{31} & Q_{33} & 0 & 0 & 0
		\end{pmatrix}$\tabularnewline\tabularnewline [1ex]
		$6'$, $\bar{6}'$, $\bar{6}'/m'$ & 
		$\begin{pmatrix}
			Q_{11} & -Q_{11} & 0 & 0 & 0 & -2Q_{22}\\
			-Q_{22} & Q_{22} & 0 & 0 & 0 & -2Q_{11}\\
			0 & 0 & 0 & 0 & 0 & 0
		\end{pmatrix}$\tabularnewline\tabularnewline [1ex]
		$6'22'$, $6'mm'$, $\bar{6}'m'2$, $\bar{6}'m2'$, $6'/m'mm'$ & 
		$\begin{pmatrix}
			Q_{11} & -Q_{11} & 0 & 0 & 0 & 0\\
			0 & 0 & 0 & 0 & 0 & -2Q_{11}\\
			0 & 0 & 0 & 0 & 0 & 0
		\end{pmatrix}$\tabularnewline\tabularnewline [1ex]
		$23$, $m\bar{3}$, $4’32'$, $\bar{4}’3m’$, $m\bar{3}m’$ & 
		$\begin{pmatrix}
			0 & 0 & 0 & Q_{14} & 0 & 0\\
			0 & 0 & 0 & 0 & Q_{14} & 0\\
			0 & 0 & 0 & 0 & 0 & Q_{14}
		\end{pmatrix}$\tabularnewline\tabularnewline [1ex]
		all other MPGs & 
		$\begin{pmatrix}
			0 & 0 & 0 & 0 & 0 & 0\\
			0 & 0 & 0 & 0 & 0 & 0\\
			0 & 0 & 0 & 0 & 0 & 0
		\end{pmatrix}$\tabularnewline\tabularnewline [0.5ex]
	\end{longtable*}
	\endgroup	
			
	\subsection{General transformation properties for Berry curvature multipoles in 3D space}
	In this subsection, we show the general transformation rules for Berry curvature multipoles and their explicit forms under certain MPG symmetries.
	
	First of all, we introduce the Jahn notation \cite{S_Jahn1949note} to describe the transformation properties of Berry curvature multipoles. A 3D polar vector is denoted by $V$, and $V^{m} = V \times V \times… \times V$ denotes a rank-$m$ tensor. The symbols $[ \enspace ]$ and $\{ \enspace \}$, denote the symmetrization and anti-symmetrization respectively of the tensors inside the symbol. $e$ and $a$ are rank-0 tensors that change sign under spatial inversion $\mathcal{I}$ and time-reversal symmetry $\mathcal{T}$, respectively. For example, the Berry curvature monopole is a pseudovector, and at the same time odd under $\mathcal{T}$,  thus transforms as $aeV$. The Berry curvature dipole is a rank-2 pseudotensor, and is even under $\mathcal{T}$, thus transforms as $eV^{2}$. For Berry curvature quadrupole, as analyzed in the previous sections, it transforms as a rank-3 pseudotensor, and is odd under $\mathcal{T}$. Furthermore, it is symmetric with respect to the first two indices which are associated with the partial derivatives, thus is of the type $ae[V^{2}]V$.
	
	Similarly, the Berry curvature hexapole $H_{\alpha\beta\gamma\delta} = \int_{\bm{k}}f_{0}\partial_{\alpha}\partial_{\beta}\partial_{\gamma}\Omega_{\delta}$ is a rank-4 pseudotensor and even under $\mathcal{T}$. A symmetry operation $\Lambda$ will impose constraints on the form of the hexapole:
	\begin{equation}
		H_{\alpha\beta\gamma\delta} =\mathrm{det}(\Lambda) \Lambda_{\alpha\alpha'}\Lambda_{\beta\beta'}\Lambda_{\gamma\gamma'}\Lambda_{\delta\delta'}H_{\alpha'\beta'\gamma'\delta'}.
	\end{equation}
	Furthermore, the hexapole is symmetric with respect to all the first three indices which are associated with the partial derivatives, and therefore transforms as $e[V^{3}]V$.
	
	In general, the $n$-th moment of Berry curvature, which can be defined as
	\begin{equation}
		P_{\alpha_{1}\alpha_{2}\dots\alpha_{n}\beta}=\int_{\bm{k}}f_{0}\partial_{\alpha_{1}}\partial_{\alpha_{2}}\dots\partial_{\alpha_{n}}\Omega_{\beta},
	\end{equation}
	transforms as a rank-($n+1$) pseudotensor, with the first $n$ indices to be symmetric. It is even under $\mathcal{T}$ for odd number of $n$, and odd under $\mathcal{T}$ for even number of $n$, thus is of the type $a^{n+1}e[V^{n}]V$.
	
	Once the transformation properties of the Berry curvature multipoles are understood, their explicit forms under certain MPGs can be obtained with the Bilbao Crystallographic Server \cite{Gallego2019automatic}. We have identified the leading-order Berry curvature moments for all the 122 3D MPGs, as tabulated in Table \ref{table:02}.
	
	\subsection{Symmetry properties for Berry curvature multipoles in 2D space}
	In this subsection, we first show that there are $n+1$ independent components for the $n$-th moment of Berry curvature in 2D space. By linear combination, they form the eigenvectors of the angular momentum operator, with quantum numbers $\pm n$, $\pm (n-2)$, $\cdots$. Apart from the zero angular momentum components, such as the monopole and the trace of quadrupole, all the other components are forced to vanish under a $p$-fold rotational symmetry with $p > n$. Based on the symmetry analysis, we further determine the leading-order Berry curvature moments for all the 31 2D MPGs.
	
	The zeroth moment of Berry curvature is the Berry curvature monopole $M=\int_{\bm{k}}f_{0}\Omega$, which has only one component as the Berry curvature becomes a pseudoscalar in 2D space. The first moment is the Berry curvature dipole $D_{\alpha}=\int_{\bm{k}}f_{0}\partial_{\alpha}\Omega$, which transforms as a pseudovector. The two components can be rearranged as $D_{\pm1}=D_{x}\pm iD_{y}=\int_{\bm{k}}f_{0}\partial_{\pm}\Omega$ with $\partial_{\pm}=\partial_{x}\pm i\partial_{y}$, which are the eigenvectors of the angular momentum operator (the generator of rotations): $\hat{L_{z}}D_{\pm1}=\pm\hbar D_{\pm1}$, with the angular moment quantum numbers $m=\pm1$. Under the rotation operator $\hat{R_z}(\theta)=\exp(-i\frac{\hat{L_{z}}}{\hbar}\theta)$, they will acquire phase factors respectively: $\hat{R_z}(\theta)D_{\pm1}=e^{\mp i\theta}D_{\pm1}$. For a $p$-fold rotational symmetry $\theta=2\pi/p$, it imposes a constraint: $e^{\mp i2\pi/p}D_{\pm1}=D_{\pm1}$, which forces the dipole to vanish since $e^{\mp i2\pi/p}\neq1$. This was first pointed out by Sodemann and Fu \cite{Sodemann2015quantum}, and we will generalize it to higher-order Berry curvature moments as shown below. Furthermore, a mirror symmetry $M_{x}$ will force the $y$ component to vanish, and the dipole will be perpendicular to the mirror plane.
	
	The Berry curvature quadrupole $Q_{\alpha\beta}=\int_{\bm{k}}f_{0}\partial_{\alpha}\partial_{\beta}\Omega$ has three independent components $Q_{xx}$, $Q_{xy}$ and $Q_{yy}$, which can be rearranged to form the eigenvectors of the angular momentum operator: $Q_{+2}=\int_{\bm{k}}f_{0}\partial_{+}^2\Omega$, $Q_{0}=\int_{\bm{k}}f_{0}\partial_{+}\partial_{-}\Omega$, $Q_{-2}=\int_{\bm{k}}f_{0}\partial_{-}^2\Omega$. $Q_{0}$ is the trace of the quadrupole, which transforms the same as the monopole $M$, and the traceless components $Q_{\pm2}$ have angular momentum $m=\pm2$ respectively. Under a $p$-fold rotation, $\hat{R_z}(\frac{2\pi}{p})Q_{m}=e^{-2i\pi m/p}Q_{m}$, which imposes a constraint on the quadrupole: $(e^{-2i\pi m/p}-1)Q_{m}=0$. For any $p>|m|$, i.e., any rotational axis with order higher than $|m|=2$, the traceless part $Q_{\pm2}$ of quadrupole is forced to vanish as $e^{-2i\pi m/p} \neq 1$. Next, considering a mirror symmetry, which for simplicity can always be written as $M_{x}$ by choosing the $x$-axis to be perpendicular to the mirror plane. The quadrupole transforms as $M_{x}Q_{m}=-Q_{-m}$, therefore the component $Q_{2}+Q_{-2}$ is odd under $M_{x}$ thus forced to vanish, while $Q_{2}-Q_{-2}$ is even under $M_{x}$ and can still be finite.
	
	It can be further generalized to the higher-order Berry curvature moments. For the $n$-th Berry curvature moment $P_{\alpha_{1}\alpha_{2}\cdots\alpha_{n}\beta}=\int_{\bm{k}}f_{0}\partial_{\alpha_{1}}\partial_{\alpha_{2}}\cdots\partial_{\alpha_{n}}\Omega_{\beta}$, there are $n+1$ independent components: $\int_{\bm{k}}f_{0}(\partial_{x})^{l}(\partial_{y})^{n-l}\Omega$, with $l=0,1,\cdots,n$. By linear combination, they can construct the eigenvectors of the angular momentum operator:
	\begin{equation}
		P_{n-2l}^{n}=\int_{\bm{k}}f_{0}\partial_{+}^{n-l}\partial_{-}^{l}\Omega,
	\end{equation}
	where the lower-index indicates the angular momentum, and the upper-index labels the order of the Berry curvature moment. Explicitly, they are $P_{n}^{n}=\int_{\bm{k}}f_{0}\partial_{+}^n\Omega$, $P_{n-2}^{n}=\int_{\bm{k}}f_{0}\partial_{+}^{n-1}\partial_{-}\Omega$, $\cdots$, $P_{-n}^{n}=\int_{\bm{k}}f_{0}\partial_{-}^n\Omega$, with quantum numbers $n$, $n-2$, $\cdots$, $-n$.
	%	All the components $P_{m}^{n}$ with $-n+2 \leq m \leq n-2$ have the same symmetry property as the lower-order moments, thus cannot be the leading-order moment. Under a p-fold rotation, $\hat{R_z}(\frac{2\pi}{p})P_{\pm n}^{n}=e^{\mp 2i\pi n/p}P_{\pm n}^{n}$, which requires $(e^{\mp 2i\pi n/p}-1)P_{\pm n}^{n}=0$. 
	Under a $p$-fold rotation, $\hat{R_z}(\frac{2\pi}{p})P_{m}^{n}=e^{- 2i\pi m/p}P_{m}^{n}$, which requires
	\begin{equation}
		(e^{- 2i\pi m/p}-1)P_{m}^{n}=0.
	\end{equation}
	For any rotational axis with order $p$ higher than the order $n$ of the Berry curvature moment, $e^{- 2i\pi m/p} \neq 1$ when $m\neq0$, thus $P_{m}^{n}$ ($m\neq0$) will be forced to vanish. Furthermore, when the mirror symmetry $M_{x}$ is present, the Berry curvature moments transform as
	\begin{equation}
		M_{x}P_{m}^{n}=(-1)^{n+1}P_{-m}^{n},
	\end{equation}
	therefore the component $P_{m}^{n}-(-1)^{n+1}P_{-m}^{n}$ is odd under $M_{x}$ thus forced to vanish, while $P_{m}^{n}+(-1)^{n+1}P_{-m}^{n}$ is even under $M_{x}$ and can still be finite.
	
	Special attention should be paid to the spatial symmetries combined with time-reversal symmetry, namely the $C_{2}\mathcal{T}$, $C_{4}\mathcal{T}$, $C_{6}\mathcal{T}$ and $M_{x}\mathcal{T}$ symmetries. In general, $p$-fold rotation combined with $\mathcal{T}$ imposes the constraint on the Berry curvature moment: $(-1)^{n+1}e^{- 2i\pi m/p}P_{m}^{n}=P_{m}^{n}$, where an additional factor $(-1)^{n+1}$ is acquired due to the $\mathcal{T}$ symmetry. Specifically, $C_{2}\mathcal{T}=(C_{6}\mathcal{T})^3$ forces the Berry curvature to vanish in the entire BZ. $C_{4}\mathcal{T}$ forces both the monopole and dipole to vanish. For quadrupole components $Q_{\pm2}$, the $C_{4}\mathcal{T}$ symmetry requires $\hat{C_4}\hat{\mathcal{T}}Q_{m}=-e^{-2i\pi m/p}Q_{m}=Q_{m}$. Since $-e^{-2i\pi m/p}=1$ for $p=4$ and $m=\pm2$, the above condition is always satisfied and $C_{4}\mathcal{T}$ has no constraint on $Q_{\pm2}$, thus a finite quadrupole is allowed. Finally, for the $M_{x}\mathcal{T}$ symmetry, the Berry curvature moments transform as $\hat{M_{x}}\hat{\mathcal{T}}P_{m}^{n}=P_{-m}^{n}$, therefore the component $P_{m}^{n}-P_{-m}^{n}$ is odd under $M_{x}\mathcal{T}$ thus forced to vanish, while $P_{m}^{n}+P_{-m}^{n}$ is even under $M_{x}\mathcal{T}$ and can still be finite.
	
	Based on the previous symmetry analysis, we can determine the leading-order Berry curvature moments in all the 2D MPGs. Out of the 31 MPGs in 2D space, 10 of them respect the $C_{2}\mathcal{T}$ symmetry, which forces the Berry curvature to vanish in the entire BZ.  Among the remaining 21 MPGs, 10 of them break both the time-reversal and mirror symmetries, therefore a non-vanishing monopole is allowed. $11'$, $m$ and $m1'$ MPGs respect at most a mirror symmetry, therefore the Berry curvature dipole is allowed. The remaining 8 MPGs are $2mm$, $4'$, $4'm'm$, $3m$, $31'$, $3m1'$, $4mm$ and $6mm$. For Berry curvature quadrupole which is the second moment of Berry curvature, any rotational symmetry with order higher than 2 will force $Q_{\pm2}$ to vanish. Therefore it can only be the leading-order moment in $2mm$, $4'$, $4'm'm$. Similarly, the Berry curvature hexapole vanishes under any rotational symmetry with order higher than 3, thus can only be the leading-order moment in $3m$, $31'$, $3m1'$. Berry curvature octopole is the leading-order moment in $4mm$, and the 12-pole is the leading-order moment in $6mm$.
	
	\renewcommand{\theequation}{D-\arabic{equation}}
	\renewcommand\thefigure{D-\arabic{figure}} 
	\setcounter{equation}{0}  
	\setcounter{figure}{0}
	\section{NLAH effect in quantum anomalous Hall materials}\label{AppendixD}

	In this appendix, we study the third-order NLAH effect induced by Berry curvature quadrupole in quantum anomalous Hall (QAH) materials. In general, the trace of quadrupole $Q_{\alpha\alpha\gamma}$ has the same symmetry property as the monopole $\int_{\bm{k}}f_{0}\Omega_{\gamma}$, thus is finite in all materials which exhibit anomalous Hall effect. In particular, QAH materials are good platforms to study the NLAH effect due to their nontrivial topological nature. Here, we will first use a general model Hamiltonian to show that sizable Berry curvature quadrupole can be achieved in QAH materials when the chemical potential is tuned to the band edge, which can possibly describe the states near the QAH phase in few-layer MnBi$_2$Te$_4$ \cite{Deng2020quantum, Fu2020exchange}. Then we will use the spin and valley polarized continuum model \cite{S_Liu2020anomalous} to describe the QAH state near 3/4 filling \cite{Sharpe2019emergent, Serlin2020intrinsic} in twisted bilayer graphene aligned with boron nitride substrate. Giant Berry curvature quadrupoles $\sim 4000 \; \mathrm{\AA^{2}}$ can be achieved, when the chemical potential is gated to the band edges.
	
	\subsection{Berry curvature quadrupole with general model Hamiltonian}
	In this subsection, we use the model Hamiltonian \cite{S_Qi2006topological, S_Liu2008quantum} to describe a QAH state:
	\begin{equation}\label{eq:QAH_Hamiltonian}
		\mathcal{H}(\bm{k}) = \epsilon(k) + v (k_{y}\sigma_{x} - k_{x}\sigma_{y}) + M(\bm{k}) \sigma_{z},
	\end{equation}	
	where $\epsilon(k)=tk^2$ is the parabolic background and $M(\bm{k})=M_0-M_2 k^2$ is the mass term, $k=|\bm{k}|$, and $\bm{\sigma}$ denotes the Pauli matrices. The system is in the QAH phase when $M_0M_2>0$.
	
\begin{figure}
		\centering
		\includegraphics[width=3.5in]{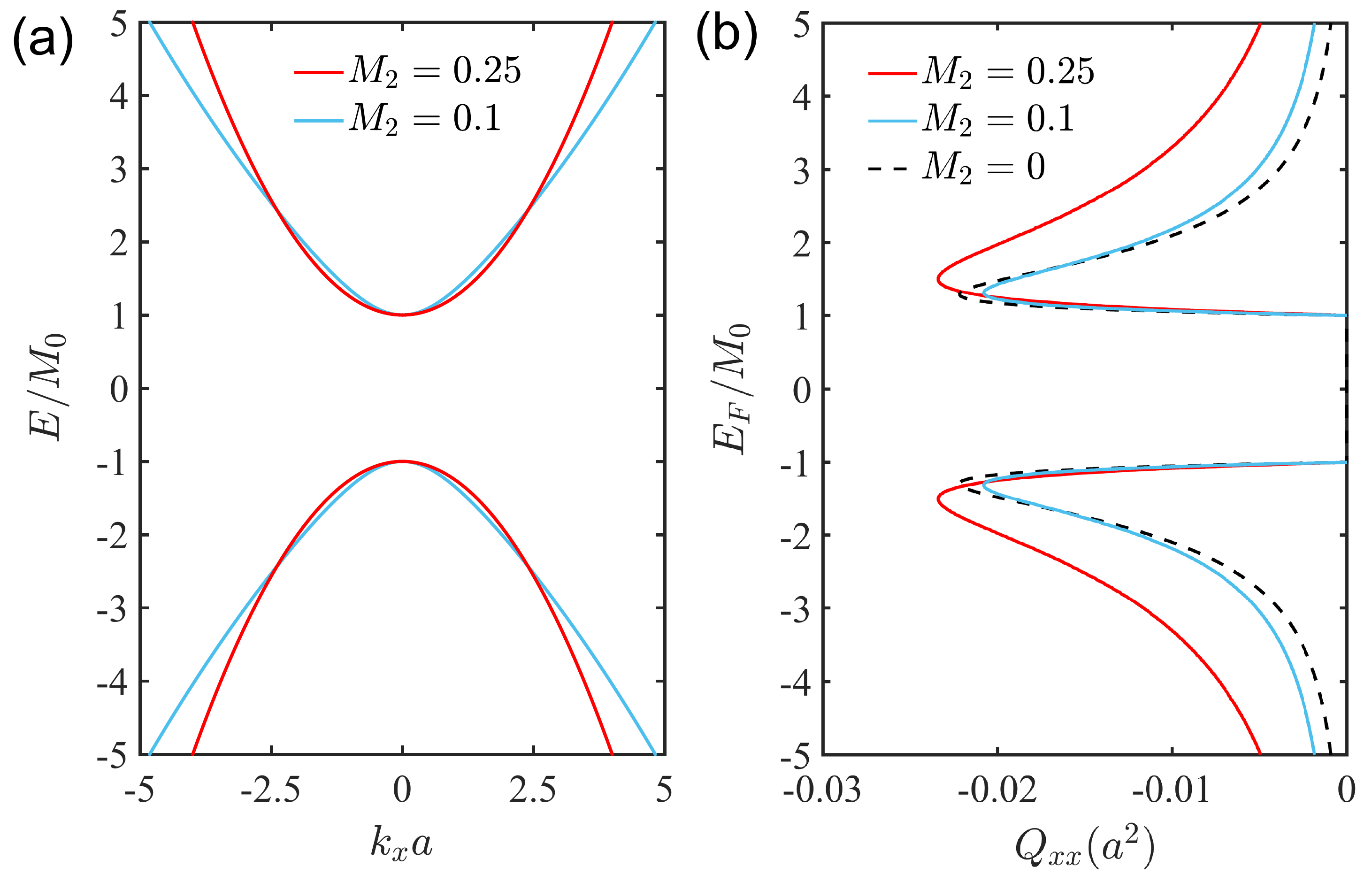}
		\caption{(a) Band structures for the QAH model in Eq. \ref{eq:QAH_Hamiltonian}, with different $M_2$ parameters.	The other parameters are $M_0 = 1$, $v = 1$, $\epsilon(k) = 0$ and the length scale $a=v/M_0$. (b) Gate dependence of the quadrupole at temperature $T = 0$. The solid dashed line is given by the analytic result in Eq. \ref{eq:BCQ_Dirac} for  $M_2=0$.}
		\label{FIGS01}
	\end{figure}

	The energy spectra of the two bands are: $E_{\pm}(\bm{k}) = t k ^ 2 \pm |\bm{d}(\bm{k})|$,  where $\pm$ denote the conduction and valence bands respectively, and $\bm{d}(\bm{k}) = [vk_{y}, -vk_{x}, M(\bm{k})]$.
	The band structures for different parameters $M_2$ are shown in Fig. \ref{FIGS01}(a).
	
	The Berry curvatures of the two bands can also be calculated as:
	\begin{equation}
		\Omega_{\pm}(\bm{k}) = \pm \frac{1}{2} \hat{\bm{d}} \cdot (\partial_{x} \hat{\bm{d}} \times \partial_{y} \hat{\bm{d}}) = \pm \frac{v^2 (M_0+M_2 k^2)}{2 |\bm{d}(\bm{k})| ^ 3},
	\end{equation}
	and the Chern numbers of the two bands are
	\begin{equation}
		\mathcal{C}_{\pm}=-2\pi\int_{\bm{k}}\Omega_{\pm}=\mp \frac{1}{2}[\mathrm{sgn}(M_0)+\mathrm{sgn}(M_2)].
	\end{equation}
	
	The model has continuous rotational symmetry, which requires $Q_{\pm2}=0$, or equivalently
	\begin{eqnarray}
		Q_{xx} &=& Q_{yy},\\
		Q_{xy} &=& 0.
	\end{eqnarray}
	The component $Q_{xx}$ is obtained numerically for different parameters $M_2$, as shown in solid lines in Fig. \ref{FIGS01}(b), which exhibits two peaks near the two band edges.
	
	In order to understand the behavior of the Berry curvature quadrupoles near band edges, we focus on the states near $k=0$, where the $M_2$-term can be neglected. For simplicity, we also neglect the $\epsilon(k)$ term which does not change the Berry curvature, and then the QAH Hamiltonian is reduced to a massive Dirac Hamiltonian. The energy dispersion is approximated as $E_{\pm}(\bm{k}) \approx \pm \sqrt{v^2k^2+M_{0}^{2}}$, and the Berry curvature $\Omega_{\pm}(\bm{k}) \approx \pm \frac{v^2 M_0}{2 (v^2k^2+M_{0}^{2})^{3/2}}$. By taking the derivative, we get $\partial_{x}E_{\pm}(\bm{k}) = \pm \frac{v^2k_x}{\sqrt{v^2k^2+M_{0}^{2}}}$ and $\partial_{x}\Omega_{\pm}(\bm{k}) = \mp \frac{3v^4 M_0 k_x}{2 (v^2k^2+M_{0}^{2})^{5/2}}$. The two bands have the same Berry curvature quadrupole, and the quadrupole of the conduction band can be calculated at zero temperature as:
	\begin{eqnarray}\label{eq:BCQ_Dirac}\nonumber
		Q_{xx} &=& \int_{\bm{k}} \partial_{x}E_{+} \partial_{x}\Omega_{+} \delta(E_{+}-\mu)\\\nonumber
		%		&=& -\frac{3v^6M_0}{2(2\pi)^2}\iint \frac{k^2\cos^2\varphi}{E_{+}^6} \delta(E_{+}-\mu) kdkd\varphi\\\nonumber
		&=& -\frac{3v^2M_0}{8\pi}\int \frac{E_{+}^2-M_{0}^2}{E_{+}^5} \delta(E_{+}-\mu) dE_{+}\\
		&=& -\frac{3v^2M_0}{8\pi} \frac{\mu^2-M_{0}^2}{\mu^5}.
	\end{eqnarray}
	The result of Eq. \ref{eq:BCQ_Dirac} is depicted in dashed line in Fig. \ref{FIGS01}(b), which can help to understand the behavior of quadrupole in the QAH model near the band edges.
	
	\subsection{NLAH effect in twisted bilayer graphene near 3/4 filling}
	
	In this subsection, we study the Berry curvature quadrupole near the QAH state of TBG near 3/4 filling \cite{Sharpe2019emergent, Serlin2020intrinsic}, with the spin and valley polarized continuum model \cite{S_Liu2020anomalous}. Giant Berry curvature quadrupoles $\sim 4000 \; \mathrm{\AA^{2}}$ can be achieved, when the chemical potential is gated to the band edges.
	%	Giant Berry curvature quadrupoles $\sim 4000 \; \mathrm{\AA^{2}}$ are found near the band edges, and the corresponding NLAH voltage is estimated.
	
	At a small twist angle $\theta$, the low-energy physics of TBG can be described by the continuum model formed by Dirac fermions in each layer \cite{S_Bistritzer2011moire}. In the layer basis, the effective Hamiltonian for valley $\xi = \pm 1$ can be written as
	\begin{equation}
		H_{0,\xi}=\begin{pmatrix}
			H_{b,\xi}(\bm{k}) & T_{\xi}(\bm{r})\\
			T_{\xi}^{\dagger}(\bm{r}) & H_{t,\xi}(\bm{k})
		\end{pmatrix}
		\label{eq:continumm},
	\end{equation}
	where $t$ ($b$) labels the top (bottom) layer, which is rotated by $+(-) \frac{\theta}{2}$ around the $z$-axis. The top layer is described by a massless Dirac Hamiltonian:
	\begin{equation}
		H_{t, \xi}(\bm{k}) = -\hbar v_{F} \hat R_{- \frac{\theta}{2}} (\bm{k}-\bm{K}_{t, \xi}) \cdot (\xi \sigma_x, \sigma_y),
	\end{equation}
	where $ \hbar v_{F} = 5.253 \; \mathrm{eV} \cdot \mathrm{\AA}$ is the original Fermi velocity, $\hat{R}$ is the rotation operator, and $\bm{\sigma}$ denotes the Pauli matrices acting on the AB sublattice space. $\bm{K}_{t/b, \xi}=\xi|K|(\frac{\sqrt{3}}{2}, \mp \frac{1}{2})$ are the BZ corners, with the magnitude $|K| = \frac{8\pi}{3a} \sin \frac{\theta}{2}$ and $ a = 2.46 \; \mathrm{\AA} $ is the graphene lattice constant.	
	
	The bottom layer Hamiltonian contains a mass term induced by the hexagonal boron nitride substrate:
	\begin{equation}
		H_{b, \xi}(\bm{k}) = -\hbar v_{F} \hat R_{\frac{\theta}{2}} (\bm{k}-\bm{K}_{b, \xi}) \cdot (\xi \sigma_x, \sigma_y) + \Delta \sigma_z,
	\end{equation}
	with $\Delta = 17$ meV \cite{S_Kim2018accurate}. The interlayer hopping is
	\begin{eqnarray}\nonumber
		T_{\xi}(\bm{r}) &=&
		\begin{pmatrix} 
			u & u' \\
			u' & u 
		\end{pmatrix}
		+
		\begin{pmatrix}
			u & u' e^{-i\xi\phi} \\
			u' e^{i\xi\phi} & u 
		\end{pmatrix}
		e^{i \xi \bm{G_{1}}\cdot\bm{r}}\\
		&&+
		\begin{pmatrix} 
			u & u' e^{i\xi\phi} \\
			u' e^{-i\xi\phi} & u
		\end{pmatrix}
		e^{i \xi (\bm{G_{1}} + \bm{G_{2}}) \cdot\bm{r}},
	\end{eqnarray}
	with $\phi = 2 \pi / 3$. $\bm{G_{1}}$ and $\bm{G_{2}}$ are the reciprocal lattice vectors of the moir\'e superlattice, which are chosen as $\bm{G_{1}}=|G|(-\frac{1}{2}, -\frac{\sqrt{3}}{2})$, $\bm{G_{2}}=|G|(1, 0)$,	with the magnitude $|G| = \sqrt{3}|K|$.	The parameters $u = 79.7$ meV and $u' = 97.5$ meV are adopted from Ref.~\cite{S_Koshino2018maximally}, which has taken the effect of lattice corrugation into consideration.
	
	In order to describe the QAH state near 3/4 filling, where the valley and spin degeneracies are lifted by interactions, we adopt the valley and spin polarized Hamiltonian from Ref. \cite{S_Liu2020anomalous}:
	\begin{equation}
		H_{\xi,s}=H_{0,\xi}+\xi E_v+sE_s,
	\end{equation}
	where $E_v$ and $E_s$ are the valley and spin splittings respectively, and $s= \pm 1$ is the spin index.
	
\begin{figure}
		\centering
		\includegraphics[width=3.5in]{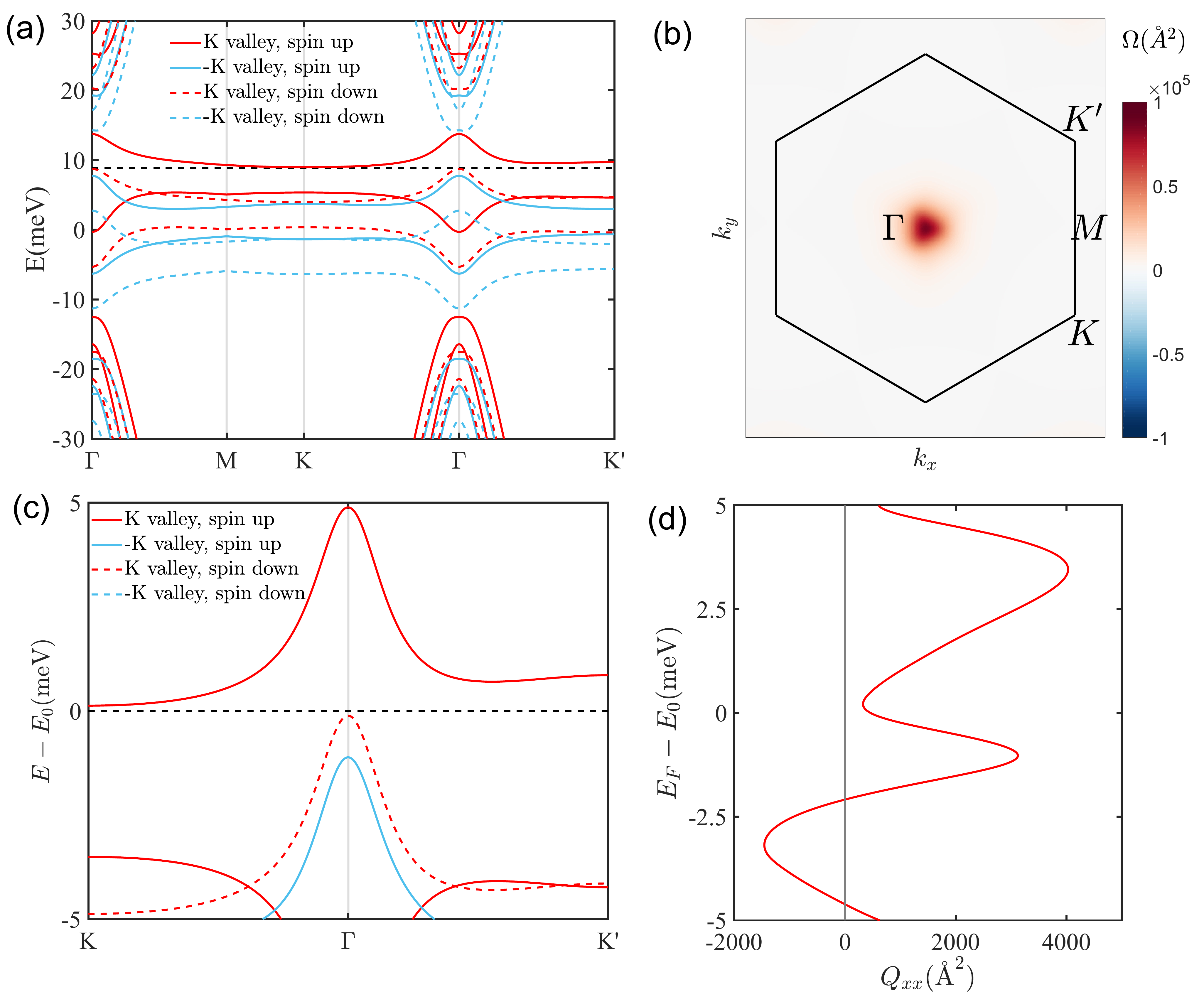}
		\caption{(a) Band structures for valley and spin polarized twisted bilayer graphene aligned with hexagonal boron nitride at $\theta=1.05\degree$, with valley splitting $E_v=3 \; \mathrm{meV}$ and spin splitting $E_s=2.5 \; \mathrm{meV}$. The high symmetry points are marked in the moir\'e BZ in Panel (b), where $K=K_{t, +}=K_{b, -}$ and $K'=K_{b, +}=K_{t, -}$. (b) Berry curvature distribution of the conduction band at $K$ valley with $s=\pm1$. (c) Enlarged energy spectra near 3/4 filling, with the origin of the energy axis shifted to $E_0=8.85 \; \mathrm{meV}$. (d) Gate dependence of the quadrupole $Q_{xx}$ near 3/4 filling, with the temperature $T = 2 \; \mathrm{K}$.}
		\label{FIGS02}
	\end{figure}

	The band structures at the magic angle $\theta=1.05\degree$ are shown in Fig. \ref{FIGS02}(a),	where the red and blue lines denote the band structures at $K$ and $-K$ valleys respectively, while solid and dashed lines represent the states with spins up and down. The black dashed line indicates the energy $E_0=8.85 \; \mathrm{meV}$ which corresponds to the 3/4 filling, and the high symmetry points are marked in the moir\'e BZ in Fig. \ref{FIGS02}(b). The enlarged energy spectra are shown in Fig. \ref{FIGS02}(c), with the origin of the energy axis shifted to $E_0$ which corresponds to 3/4 filling.
	
	The conduction bands at $K$ valley with spin up and down have the same Chern number $\mathcal{C}=-1$, and the same Berry curvature distribution as shown in Fig. \ref{FIGS02}(b), while the bands at $-K$ valley have the Chern number $\mathcal{C}=1$. At 3/4 filling, three of the conduction bands are filled while one band is left empty, leading to the QAH phase with Chern number $\mathcal{C}=1$.
	
	The pristine TBG respects the $D_6$ symmetry, which is reduced to $C_3$ when it is aligned with the hexagonal boron nitride substrate. The $C_3$ symmetry requires $Q_{xx} = Q_{yy}$ and	$Q_{xy} = 0$, and the gate dependence of $Q_{xx}$ component is shown in Fig. \ref{FIGS02}(d). When the chemical potential is gated to the band edges, giant Berry curvature quadrupoles $\sim 4000 \; \mathrm{\AA}^2$ can be achieved.
	
	\renewcommand{\theequation}{E-\arabic{equation}}
	\renewcommand\thefigure{E-\arabic{figure}} 
	\setcounter{equation}{0}  
	\setcounter{figure}{0}
	\section{Angular dependence of the NLAH responses in 2D space}\label{AppendixE}

	In this appendix, we discuss the angular dependence of the NLAH responses induced by Berry curvature multipoles in 2D space for three purposes. First of all, by applying the current in a general direction, we see how the multipole components contribute to the NLAH voltage so that their physical meanings can be understood. In general, the NLAH voltage induced by the Berry curvature moment with angular momentum quantum number $m$ has $m$-fold angular dependence. Second, due to the unique angular dependence of the NLAH voltage, it can be a characteristic signature to identify the NLAH response. Third, the angular dependence can also be used to distinguish the NLAH responses related to the anti-symmetric part of the conductivity tensor from the Drude-like contributions which are related to the symmetric part of the conductivity tensor.
	
	\subsection{Third-order NLAH response induced by Berry curvature quadrupole}
	
	In the DC limit $\omega \ll \frac{1}{\tau}$, The NLAH and Drude-like contribution to the third harmonic generation can be simplified as:
	\begin{eqnarray}
		\sigma^{(3),H}_{\mu\alpha\beta\gamma}(3\omega)&=&-\frac{e^{4}\tau^{2}}{4\hbar^{3}}\int_{\bm{k}}f_{0}\partial_{\beta}\partial_{\gamma}\mathcal{F}_{\mu\alpha},\\
		\sigma^{(3),D}_{\mu\alpha\beta\gamma}(3\omega)&=&\frac{e^{4}\tau^{3}}{4\hbar^{4}}\int_{\bm{k}}f_{0}\partial_{\mu}\partial_{\alpha}\partial_{\beta}\partial_{\gamma}\varepsilon_{\bm{k}}.
	\end{eqnarray}
	
	A generic electric field $\bm{E}(\omega)=(E_{x}, E_{y})$ will generate a third-order NLAH current $\bm{j}^{H}(3\omega)=(j_{x}^{H}, j_{y}^{H})$, with
	\begin{eqnarray}\nonumber
		j_{x}^{H}&=&-\frac{e^{4}\tau^{2}}{4\hbar^{3}}(Q_{xx}E_{x}^{2}E_{y}+2Q_{xy}E_{x}E_{y}^{2}+Q_{yy}E_{y}^{3}),\\\label{eq:Hall_x}
		&&\\\label{eq:Hall_y}
		j_{y}^{H}&=&\frac{e^{4}\tau^{2}}{4\hbar^{3}}(Q_{xx}E_{x}^{3}+2Q_{xy}E_{x}^{2}E_{y}+Q_{yy}E_{x}E_{y}^{2}),
	\end{eqnarray}
	as well as a Drude-like contribution $\bm{j}^{D}(3\omega)=(j_{x}^{D}, j_{y}^{D})$, with
	\begin{eqnarray}\nonumber
		j_{x}^{D}&=&\sigma^{D}_{xxxx}E_{x}^{3}+3\sigma^{D}_{xxxy}E_{x}^{2}E_{y}+3\sigma^{D}_{xxyy}E_{x}E_{y}^{2}+\sigma^{D}_{xyyy}E_{y}^{3},\\\label{eq:Drude_x}
		&&\\\nonumber
		j_{y}^{D}&=&\sigma^{D}_{xxxy}E_{x}^{3}+3\sigma^{D}_{xxyy}E_{x}^{2}E_{y}+3\sigma^{D}_{xyyy}E_{x}E_{y}^{2}+\sigma^{D}_{yyyy}E_{y}^{3}.\\\label{eq:Drude_y}
		&&
	\end{eqnarray}
	
	Now, we consider a general case in which a current $\bm{j}(\omega)=j(\cos\theta, \sin\theta)$ is applied at angle $\theta$ away from the $x$ direction, where $j=I/W$ is the current density and $W$ is the width in the transverse direction. For simplicity, we consider an isotropic linear resistivity $\bm{\rho}^{(1)}(\omega)= \rho_{0}
	\begin{pmatrix}
		1 & 0\\
		0 & 1
	\end{pmatrix}$,
	which can be achieved in systems with $p$-fold $(p\geq3)$ rotational symmetry. Then the corresponding applied electric field $\bm{E}(\omega) = \rho_{0}j(\cos\theta, \sin\theta)$. The third-order NLAH current induced by the electric field can be obtained by Eqs. \ref{eq:Hall_x} and \ref{eq:Hall_y} as
	\begin{eqnarray}\nonumber
		j_{\perp}^{H}(3\omega)&=& -j_{x}^{H}\sin\theta+j_{y}^{H}\cos\theta\\\nonumber
		&=&\frac{e^{4}\tau^{2}}{8\hbar^{3}}(\rho_{0}j)^{3}[(Q_{xx}+Q_{yy})\\\nonumber
		&&+(Q_{xx}-Q_{yy})\cos(2\theta)\\
		&&+2Q_{xy}\sin(2\theta)].
	\end{eqnarray}
	Or equivalently the NLAH voltage is
	\begin{eqnarray}\nonumber
		\frac{V_{\perp}^{H}(3\omega)/W}{(I/W)^3}&=&\frac{e^{4}\tau^{2}}{8\hbar^{3}}\rho_{0}^{4}[(Q_{xx}+Q_{yy})\\\nonumber
		&&+(Q_{xx}-Q_{yy})\cos(2\theta)\\
		&&+2Q_{xy}\sin(2\theta)].
	\end{eqnarray}
	The contribution from the trace of quadrupole has no angular dependence, while the contributions from the traceless components have two-fold angular dependence.
	
	The third-order Drude-like current can be obtained by Eqs. \ref{eq:Drude_x} and \ref{eq:Drude_y}, and its component along the transverse direction ($\theta+\frac{\pi}{2}$) is
	\begin{eqnarray}\nonumber
		j_{\perp}^{D}(3\omega)&=& -j_{x}^{D}\sin\theta+j_{y}^{D}\cos\theta\\\nonumber
		&=&\frac{1}{4}(\rho_{0}j)^{3}[(-\sigma^{D}_{xxxx}+\sigma^{D}_{yyyy})\sin(2\theta)\\\nonumber
		&&+2(\sigma^{D}_{xxxy}+\sigma^{D}_{xyyy})\cos(2\theta)\\\nonumber
		&&+(3\sigma^{D}_{xxyy}-\frac{1}{2}\sigma^{D}_{xxxx}-\frac{1}{2}\sigma^{D}_{yyyy})\sin(4\theta)\\
		&&+2(\sigma^{D}_{xxxy}-\sigma^{D}_{xyyy})\cos(4\theta)].
	\end{eqnarray}
	Or equivalently the corresponding voltage is
	\begin{eqnarray}\nonumber
		\frac{V_{\perp}^{D}(3\omega)/W}{(I/W)^3}&=&\frac{1}{4}\rho_{0}^{4}[(-\sigma^{D}_{xxxx}+\sigma^{D}_{yyyy})\sin(2\theta)\\\nonumber
		&&+2(\sigma^{D}_{xxxy}+\sigma^{D}_{xyyy})\cos(2\theta)\\\nonumber
		&&+(3\sigma^{D}_{xxyy}-\frac{1}{2}\sigma^{D}_{xxxx}-\frac{1}{2}\sigma^{D}_{yyyy})\sin(4\theta)\\
		&&+2(\sigma^{D}_{xxxy}-\sigma^{D}_{xyyy})\cos(4\theta)].
	\end{eqnarray}
	The first two terms contributed by $-\sigma^{D}_{xxxx}+\sigma^{D}_{yyyy}$ and 
	$\sigma^{D}_{xxxy}+\sigma^{D}_{xyyy}$ have the same symmetry properties as the anisotropic conductivity $-\sigma^{D}_{xx}+\sigma^{D}_{yy}$ and $\sigma^{D}_{xy}$ respectively, which vanish under $p$-fold $(p\geq3)$ rotational symmetry or $p$-fold rotation combined with time-reversal symmetry. The contributions from last two terms have four-fold angular dependence.
	
	As a conclusion, in isotropic 2D materials, the third-order NLAH contributions which have two-fold or no angular dependence, can be easily distinguished from the Drude-like contributions which have four-fold angular dependence.
	
	Especially, under $4'm'm$ MPG which is the case for monolayer SrMnBi$_2$, $Q_{xx}=-Q_{yy}$, $Q_{xy}=0$, $\sigma^{D}_{xxxx}=\sigma^{D}_{yyyy}$, $\sigma^{D}_{xxxy}=\sigma^{D}_{xyyy}=0$, the NLAH and the Drude-like contributions can be simplified as:
	\begin{eqnarray}	
		\frac{V_{\perp}^{H}(3\omega)/W}{(I/W)^3}&=&\frac{e^{4}\tau^{2}}{4\hbar^{3}}\rho_{0}^{4}Q_{xx}\cos(2\theta),\\
		\frac{V_{\perp}^{D}(3\omega)/W}{(I/W)^3}&=&\frac{1}{4}\rho_{0}^{4}(3\sigma^{D}_{xxyy}-\sigma^{D}_{xxxx})\sin(4\theta).
	\end{eqnarray}
	The NLAH voltage shows a two-fold angular dependence,  while in contrast the Drude-like response shows a four-fold angular dependence. In particular, when the electric current is applied along the $x$ or $y$ direction, the Drude-like contribution vanishes and the NLAH voltage optimizes.
	
	\subsection{Fourth-order NLAH response induced by Berry curvature hexapole}
	Since the Drude-like contribution to the fourth-order response vanishes in time-reversal invariant systems, we will only focus on the angular dependence of the NLAH contribution.
	
	In the DC limit $\omega \ll \frac{1}{\tau}$, The NLAH contribution to the fourth harmonic generation can be simplified as:
	\begin{equation}
		\sigma^{(4),H}_{\mu\alpha\beta\gamma\delta}(4\omega)=\frac{e^{5}\tau^{3}}{8\hbar^{4}}\int_{\bm{k}}f_{0}\partial_{\beta}\partial_{\gamma}\partial_{\delta}\mathcal{F}_{\mu\alpha}.
	\end{equation}
	
	A generic electric field $\bm{E}(\omega)=(E_{x}, E_{y})$ will generate a fourth-order NLAH current $\bm{j}^{H}(4\omega)=(j_{x}^{H}, j_{y}^{H})$, with
	\begin{eqnarray}\nonumber
		j_{x}^{H}&=&\frac{e^{5}\tau^{3}}{8\hbar^{4}}(H_{xxx}E_{x}^{3}E_{y}+3H_{xxy}E_{x}^{2}E_{y}^{2}\\
		&&+3H_{xyy}E_{x}E_{y}^{3}+H_{yyy}E_{y}^{4}),\\\nonumber
		j_{y}^{H}&=&-\frac{e^{5}\tau^{3}}{8\hbar^{4}}(H_{xxx}E_{x}^{4}+3H_{xxy}E_{x}^{3}E_{y}\\
		&&+3H_{xyy}E_{x}^{2}E_{y}^{2}+H_{yyy}E_{x}E_{y}^{3}),
	\end{eqnarray}
	which can be rewritten as
	\begin{eqnarray}\nonumber
		j_{\pm}^{H}&=&j_{x}^{H}\pm i j_{y}^{H}\\\nonumber
		&=&\mp\frac{e^{5}\tau^{3}}{64\hbar^{4}}i(H_{3}E_{-}^3+3H_{1}E_{-}^2E_{+}\\\label{eq:Hall_pm}
		&&+3H_{-1}E_{-}E_{+}^2+H_{-3}E_{+}^3)E_{\pm},
	\end{eqnarray}
	where $E_{\pm}=E_{x}\pm iE_{y}$, $H_{n-2l}=\int_{\bm{k}}f_{0}\partial_{+}^{n-l}\partial_{-}^{l}\Omega$ with $n=3$ and $l=0,1,2,3$.
	
	Now, we consider a general case in which a current $\bm{j}(\omega)=j(\cos\theta, \sin\theta)$ is applied at angle $\theta$ away from the $x$ direction, where $j=I/W$ is the current density and $W$ is the width in the transverse direction. For simplicity, we consider an isotropic linear resistivity $\bm{\rho}^{(1)}(\omega)= \rho_{0}
	\begin{pmatrix}
		1 & 0\\
		0 & 1
	\end{pmatrix}$,
	then the corresponding applied electric field $\bm{E}(\omega) = \rho_{0}j(\cos\theta, \sin\theta)$, or equivalently $E_{\pm}=\rho_{0}je^{\pm i\theta}$. The fourth-order NLAH current induced by the electric field can be obtained by Eq. \ref{eq:Hall_pm} as
	\begin{eqnarray}\nonumber
		j_{\perp}^{H}(4\omega)&=& -j_{x}^{H}\sin\theta+j_{y}^{H}\cos\theta\\\nonumber
		&=&\frac{1}{2i}(j_{+}^{H}e^{-i\theta}-j_{-}^{H}e^{i\theta})\\\nonumber
		&=&-\frac{e^{5}\tau^{3}}{64\hbar^{4}}(\rho_{0}j)^{4}(H_{3}e^{-3i\theta}+3H_{1}e^{-i\theta}\\
		&&+3H_{-1}e^{i\theta}+H_{-3}e^{3i\theta}).
	\end{eqnarray}
	Or equivalently the NLAH voltage is
	\begin{eqnarray}\nonumber
		\frac{V_{\perp}^{H}(4\omega)/W}{(I/W)^4}&=&-\frac{e^{5}\tau^{3}}{64\hbar^{4}}\rho_{0}^{5}(H_{3}e^{-3i\theta}+3H_{1}e^{-i\theta}\\
		&&+3H_{-1}e^{i\theta}+H_{-3}e^{3i\theta}).
	\end{eqnarray}
	The contributions from $H_{\pm1}$ have one-fold angular dependence, while the contributions from $H_{\pm3}$ have three-fold angular dependence.
	
	In particular, under $3m1'$ ($C_{3v}$) MPG which is the case for the surface states of topological insulators, $H_{\pm 1}=0$, $H_3=H_{-3}=4H_{xxx}$, and the NLAH response can be simplified as:
	\begin{equation}	
		\frac{V_{\perp}^{H}(4\omega)/W}{(I/W)^4}=-\frac{e^{5}\tau^{3}}{8\hbar^{4}}\rho_{0}^{5}H_{xxx}\cos(3\theta).
	\end{equation}
	The NLAH voltage shows a three-fold angular dependence, and optimizes when the electric current is applied perpendicular to the mirror plane.
	
	\bibliographystyle{apsrev4-1}
	\bibliography{Reference}

\begin{thebibliography}{61}
\bibitem [{Not()}]{Notation}%
  \BibitemOpen
  \href@noop {} {}\bibinfo {note} {Furthermore, a prime attached to a symmetry
  operation indicates that this operation is combined with the time-reversal
  operation.}\BibitemShut {Stop}%
\bibitem [{Ome()}]{OmegaTilde}%
  \BibitemOpen
  \href@noop {} {}\bibinfo {note} {For example, $\widetilde{\omega}$ represents
  $i\omega+\gamma$, $\widetilde{2\omega}$ represents $2i\omega+\gamma$ and
  $\widetilde{-\omega}$ represents $-i\omega+\gamma$. In the DC limit $\omega
  \ll \frac{1}{\tau}$, they can be simplified as $\widetilde{n\omega}
  \rightarrow \gamma = \tau^{-1}$.}\BibitemShut {Stop}%
\end{thebibliography}
	
\end{document}